\documentclass[journal]{IEEEtran}
\IEEEoverridecommandlockouts
\usepackage{cite}
\usepackage{amsmath,amssymb,amsfonts}
\usepackage{algorithmic}
\usepackage{graphicx}
\usepackage{textcomp}
\usepackage{epstopdf}
\usepackage{subfigure}
\usepackage{xcolor}
\usepackage{amsthm}
\theoremstyle{definition}
\newtheorem{definition}{Definition}[section]
\newtheorem*{remark}{Remark}
\def\BibTeX{{\rm B\kern-.05em{\sc i\kern-.025em b}\kern-.08em
    T\kern-.1667em\lower.7ex\hbox{E}\kern-.125emX}}
\begin{document}
\title{Strategic Signaling for Utility Control in Audit Games}

\author{{Jianan~Chen,
        Qin~Hu,
        and~Honglu~Jiang}
\thanks{Jianan Chen and Qin Hu (Corresponding Author) are with the Department of Computer and Information Science, Indiana University - Purdue University Indianapolis, IN, USA. Email: jc144@iu.edu, qinhu@iu.edu}       
\thanks{Honglu Jiang is with the Department of Informatics and Engineering Systems, The University of Texas Rio Grande Valley, Brownsville, TX, USA. Email: honglu.jiang@utrgv.edu}
\thanks{This work is partly supported by the US NSF under grant CNS-2105004.}}
\maketitle

\begin{abstract} 
As an effective method to protect the daily access to sensitive data against malicious attacks, the audit mechanism has been widely deployed in various practical fields. In order to examine security vulnerabilities and prevent the leakage of sensitive data in a timely manner, the database logging system usually employs an online signaling scheme to issue an alert when suspicious access is detected. Defenders can audit alerts to reduce potential damage. This interaction process between a defender and an attacker can be modeled as an audit game. In previous studies, it was found that sending real-time signals in the audit game to warn visitors can improve the benefits of the defender. However, the previous approaches usually assume perfect information of the attacker, or simply concentrate on the utility of the defender. In this paper, we introduce a brand-new zero-determinant (ZD) strategy to study the sequential audit game with online signaling, which empowers the defender to unilaterally control the utility of visitors when accessing sensitive data. In addition, an optimization scheme based on the ZD strategy is designed to effectively maximize the utility difference between the defender and the attacker. Extensive simulation results show that our proposed scheme enhances the security management and control capabilities of the defender to better handle different access requests and safeguard the system security in a cost-efficient manner.
\end{abstract}

\section{Introduction}
Since the databases of modern organizations store a large amount of private information, such as personal health and commercial secrets, their sensitivity and economic value make the databases prominent targets of malicious attacks or illegal invasions. Therefore, audit mechanisms are widely deployed, which utilize a combination of manual operations and automated methods to detect and deter attackers. Currently, the audit mechanism has become a typical method employed by many organizations with a large amount
of sensitive information, such as hospitals, banks, and search engine companies, to prevent information security attacks \cite{barth2007privacy}.

Despite the extensive employment of audit mechanisms, information leakage and illegal transactions caused by various attacks are still widespread according to a recent report \cite{news}. This concern can be even worse as some internal malicious users can abuse their authority to launch attacks. These vicious attacks from inside are less likely to be audited because they have certain privileges. To deal with these problems, modern databases are usually equipped with alarm functions in the audit mechanism to notify visitors and defenders of the potential risks during access to critical information \cite{terzi2015survey,hasan2016secure}. These alerts, which will be sent to defenders, are triggered by some specific access requests meeting predefined rules. In some audit mechanisms, users (or attackers) are granted with access permissions by defenders. And these granted permissions will be recorded in the log so that defenders can retrospectively check for any potential abuse or attack. 

Currently, researchers usually model the above audit process between the defender and the attacker as an audit game\cite{blocki2012audit,blocki2015audit,yan2018get}. To further enhance the timeliness in this process, other researchers introduce a signaling scheme working in an online manner. 
Whenever an access request triggers an alarm, the auditor will send a signal to the visitor to remind him/her that the requested data are sensitive. The behavior of sending a signal can be real-time with manual operations, or it can be automatic according to offline-setting rules. Although signaling does not substantially defend against attacks, it can help defenders discover security vulnerabilities promptly and prevent attackers from making more severe damages. In addition, the signaling step can interfere with attackers by strategically disclosing noisy information. The effectiveness of signaling has been proved in \cite{xu2015exploring}, and there are several studies \cite{blocki2012audit,dughmi2019algorithmic,yan2020warn} based on the Stackelberg game providing auditors with better strategic guidance in defending the database. In the industry, multiple medical centers and online service websites have deployed signaling schemes to protect sensitive data \cite{hedda2017evaluating}.

However, there exist two major shortcomings of the current research on signaling-based audit games. First, the widely employed Stackelberg game model usually assumes perfect information of attackers, which can be unrealistic since attackers may adopt various strategies in practice \cite{yan2020warn,sinha2018stackelberg}. Second, the existing studies focus more on the defender's interest without considering the attacker's utility, implying that the higher interest of the defender corresponds to the lower utility of the attacker\cite{terzi2015survey,blocki2015audit,laszka2017game}, which may not hold for all types of attacks.

In this paper, we model the interactions between the defender and the attacker as a sequential game, where the attacker can observe the action of the defender regarding sending signals. In this sequential game, the defender acting first will be at a disadvantage, since the attacker can make more beneficial choices after witnessing the defender's behavior, leading to enormous losses for the defender in the long run. To solve this problem, given that the defender cannot fully detect the attacker’s strategy, a brand-novel approach is employed to allow the defender to play against various attackers flexibly. More specifically, no matter what strategy the attacker employs, the defender can always deliberately set a feasible strategy of signaling and auditing to control the damages brought by the attack. Furthermore, compared with the existing methods, our proposed strategy is more in line with the real audit environment where the defender may not be able to predict the specific strategy adopted by an attacker. To achieve these goals, we employ the zero-determinant (ZD) strategy \cite{press2012iterated} to analyze the sequential audit game, which empowers the defender to unilaterally manage the utility of the attacker and even the utility difference between the defender and the attacker. By this means, we can address  the issues of the existing studies, where the perfect information of the attacker is not required, but the interest of the attacker is explicitly considered and restricted.

Our main contributions can be summarized as follows:
\begin{itemize}
\item  Considering that the audit action of the defender might be deterministic or probabilistic, we propose two different sequential games to model the interactions between the defender and the attacker, which describes the audit game in a more comprehensive manner.
\item To unilaterally control the attacker's utility, we introduce a strategy guide for the defender with the help of the extended ZD strategy, which enables the defender to set up defense strategies for a low utility of the attacker in an effective way. Besides, we reveal the critical strategy variable in utility control for the defender by analyzing the controlling gradients and value ranges.
\item For the cost-efficiency of utility control, we design an optimization scheme based on the ZD strategy to maximize the utility difference between the defender and the attacker, instead of controlling the utility of the attacker solely.
\item Through comparing with classic strategies, we evaluate the effectiveness of our proposed ZD strategy-based schemes, where the defender adopting the ZD strategy can efficiently control the utility of the attacker using various strategies, and further maximize the utility difference between the defender and the attacker.
\end{itemize}

The remainder of this paper is organized as follows. We introduce the most related work in Section II. Two game models are presented in Section III. Section IV displays how the defender uses the ZD strategy to unilaterally control the attacker's utility. Section V proposes an optimization scheme to control the utility difference between the defender and the attacker. Experimental evaluation is reported in Section VI and the whole paper is concluded in Section VII.

\section{Related Work}\label{section2}

Most of the research on audit games focus on three aspects: dealing with different types of alarms, adapting to actual database scenarios, and optimizing the expected utility of the defender.

In order to solve the challenge of handling different types of alarms, Yan et al. proposed a game-theoretic audit method which first determines the priority of different alarms, and then assigns distinct amounts of resources to alarms with resource upper limits \cite{yan2018get,yan2019database}. Schlenker et al. proposed a method to distribute appropriate alerts to security analysts for different fields \cite{schlenker2017don}. In \cite{xu2015exploring}, based on the two-stage security game framework, Xu et al. studied this problem by solving an optimization problem of Stacklberg equilibrium with a developed scalable approach.

Regarding the extension to the real-world scenario, Blocki et al. generalized the audit game model to account for multiple audit resources where each resource is restricted to audit a subset of potential violations \cite{blocki2013audit}. Korzhyk et al. designed a polynomial time algorithm for security games with multiple resources \cite{korzhyk2010complexity}. Schlenker et al. used an approach based on game theory to address alerts\cite{schlenker2017don}, which can be well extended to different database security applications. Kiral et al. analyzed the inherent role conflicts of internal audit in risk management using signal game model\cite{kiral2020resolution}.

Optimizing the expected utility of auditors can bring direct economic benefits to the database, where the related research can be divided into two categories: classic security game based and two-stage security game based. Blocki et al. first modeled the audit problem between an auditor and an auditee as a classic security game \cite{blocki2013audit}. In this case, the auditor takes a strategic action with the goal of learning an optimized resource allocation strategy to optimize the auditor's expected utility. However, other research \cite{yan2020warn} claimed that the scalability of this framework is limited since the methods in \cite{blocki2013audit,blocki2015audit} regarded alerts as targets that could be attacked, which are not easy to apply to database. Xu et al. proposed a two-stage security game framework to overcome this challenge \cite{xu2015exploring}, where the characteristic is that the defender will leak his own information and send a signal in the second stage, which can protect the target with a better performance. A subsequent work \cite{dughmi2019algorithmic} extended the advantages of signaling to Stackelberg games. This shows that the signal can also enhance the defense performance in the security game to a certain extent.

Our work is more related to optimizing the expected utility of the defender, which is usually modeled as Stackelberg games in previous studies. It is worth noting that the Stackelberg game requires complete information, which is difficult to achieve in a real audit environment. In the face of unknown strategy attacks, defenders need to respond more efficiently, which inspires this paper. Besides, previous research pay more attention to the utility of the defender, but lacked research on attackers' behaviors. We use the ZD strategy in this paper to allow the defender to have more control over the attacker's utility with unknown strategies, which has no requirement on the information completeness of audit games.

\section{Game Models} \label{Modeling}
We consider the interaction process occuring between an attacker and a defender, starting with the attacker issuing an access request for a certain type of data. Access to different types of data will trigger different types of alerts. 
After one type of alert verifies the access permission which does not necessarily ensure security, but allows the visitor to enter the database, the defender receives the alarm and chooses whether to send a signal for real-time notification. The content of the prompt can be like if the attacker continues to visit, it may be reviewed. At this point, when receiving a prompt, the attacker clearly knows that the defender has sent a signal. Next, the attacker can further choose to continue access (and carry out illegal activities) or exit directly according to whether he receives the signal. After performing this operation, the defender will decide whether to audit based on whether there is a signal sent.

\begin{figure}
\centering
\includegraphics[scale=0.45]{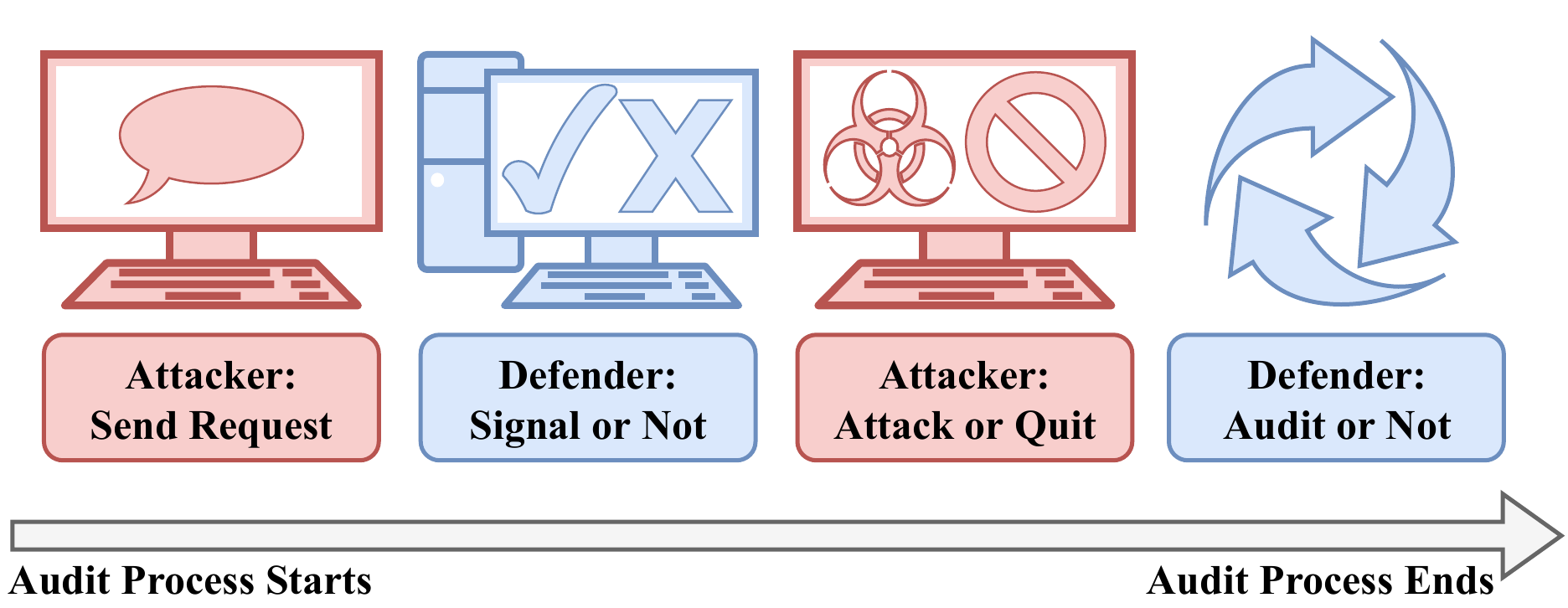}
\caption{The interaction process between the attacker and the defender.}
\label{fig:process}
\end{figure}

As shown in Fig. \ref{fig:process}, the attacker first sends a request, and then the defender chooses whether to send a signal according to his request. Next, the attacker decides whether to continue or not based on the signaling behavior of the defender. The actions of both parties are carried out strictly in order, and the previous actions of the other party can be observed. Therefore, we define the interaction process between the attacker and defender as a sequential game.

Since the attacker can access the database multiple times or different units of the database, the defender interacts with the same attacker in multiple rounds, leading to an iterative sequential game. Participants of this kind of non-zero-sum iterative sequential game may get into trouble because of the existence of dominant strategies. In practical, this is not conducive to defenders. Nash equilibrium reveals the dominant strategies of both parties in this type of game. Therefore, we study the Nash equilibrium that may exist in this game, hoping to adopt a reasonable strategy to control the attacker's utility. 

The defender will choose whether to audit or not according to the situation of sending signals after the attacker's action. The relationship between signaling and auditing can be probabilistic, where the defender audits with a certain probability. Or it could be deterministic, where the defender only audits after sending a signal. In the following context, in order to study the defender's strategy more comprehensively, we establish two models, the deterministic model and the probabilistic model.

\subsection{Deterministic Model}

In the deterministic model, the correlation between the defender's signaling and auditing is simple: for the alert of type $\eta$, if the defender sends a signal for the attacker's request, she\footnote{For the sake of distinction, we use ``she" to refer to the defender and ``he" to refer to the attacker.} will definitely audit the request; otherwise she will not. We denote the action of the defender as $d\in\{0,1\}$, where 0 represents that the defender chooses not to send a signal to the current request or audit it, while 1 indicates that the defender sends a signal to the current request and audits it. The attacker’s action is denoted as $a\in\{0,1\}$, where 0 refers to attack and 1 refers to quit without further attacks. Thus, there are four possible states of the game between the defender and the attacker, i.e., $da=(00,01,10,11)$.

\begin{figure}
\centering
\includegraphics[scale=0.45]{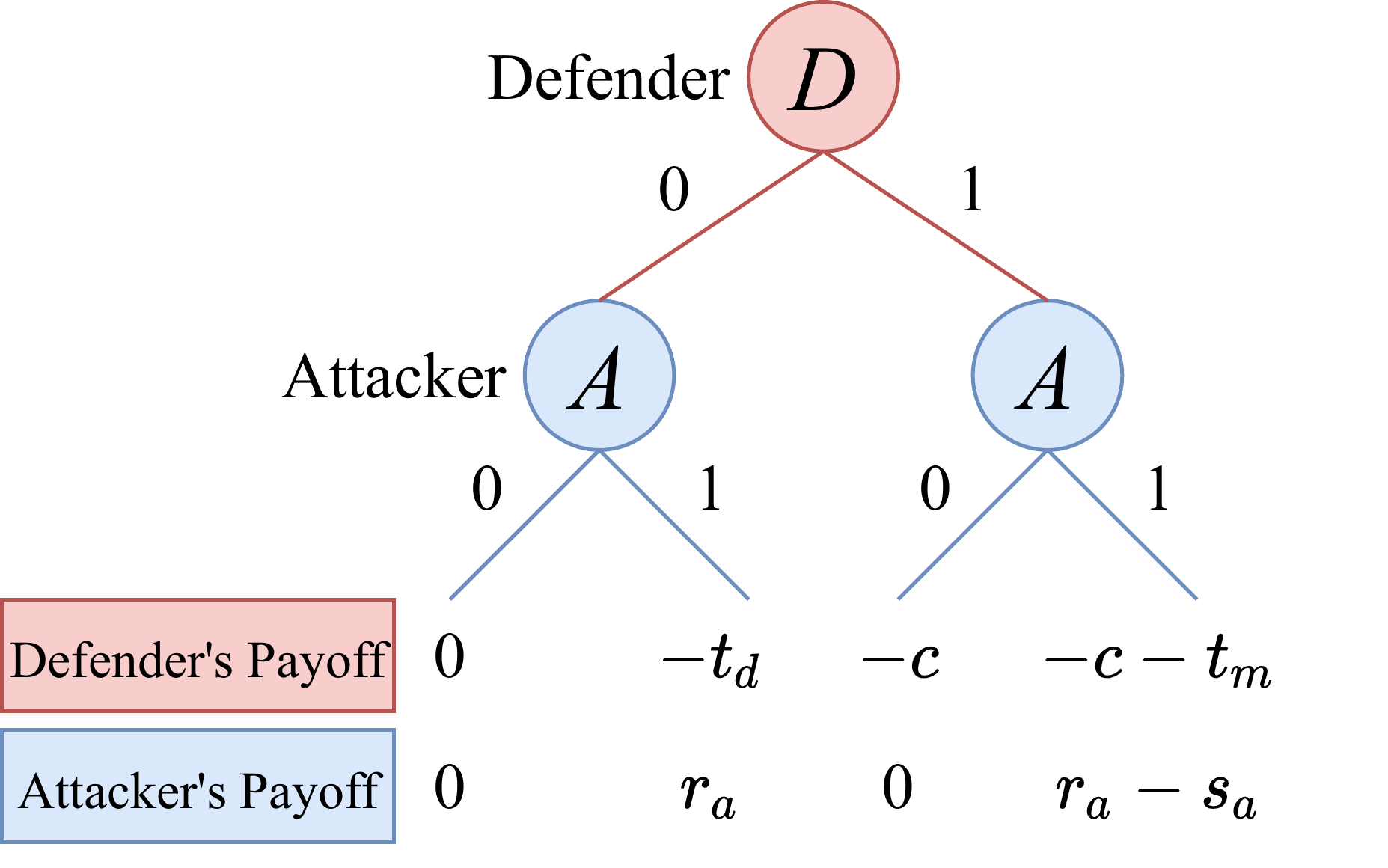}
\caption{The game tree of the deterministic model.}
\label{fig:naiveTreePayoff} 
\end{figure}

We can depict the sequential interaction process between the defender and the attacker in one round using a game tree as shown in Fig. \ref{fig:naiveTreePayoff}. In the game tree, the payoffs in four states can be calculated as follows: i) for $da=00$, as the defender chooses not to send a signal and the attacker doesn't attack, no one costs or acquires anything; ii) for $da=01$, since the defender chooses not to 
send a signal but the attacker continues to attack, the defender suffers a loss $t_d$ without auditing, while the attacker gains income $r_a$ from a successful attack; iii) for $da=10$, the defender sends a signal and audits but the attacker quits, so the defender spends $c$ as the cost of auditing while the attacker acquires nothing; iv) when $da=11$, meaning that the defender sends a signal and audits while the attacker deploys malicious attack, the defender suffers a loss $t_m$ plus the cost of audit $c$, where $t_m$ denotes the loss of being attacked but auditing timely; as for the attacker, the audit operation brings the attacker a decrease of $s_a$ on income $r_a$, where $s_a$ refers to the loss of the attacker being audited. Subsequently, we can define the payoff vectors of the defender ($D$) and the attacker ($A$) as:
\begin{gather*}
\mathbf{U}^{\eta}_D=(0,-t_d,-c,-c-t_m),\\
\mathbf{U}^{\eta}_A=(0,r_a,0,r_a-s_a).
\end{gather*}
It should be noted that $t_d, c, t_m, r_a,$ and $s_a$ are all positive. In particular, for the attacker, once the attack is successful without being caught, the benefit is large since the attacker can obtain valuable information or destroy the database. While for the defender, timely auditing after the attack or taking other repair measures, such as rollback, can only reduce the defender's loss. For the defender, auditing the attack can bring more benefit, i.e., the loss of non-auditing is larger than that of auditing. Therefore, we assume $t_d>t_m+c$. From the defender's point of view, 
she can gain from the historical data about the attacker's income  $r_a$ and the loss caused by the audit $s_a$. These two values, $r_a$ and $s_a$, help the defender to control the attacker's utility in future games.

From the perspective of the attacker being the last player to perform action, the action with the greatest benefit is 1, so he makes this choice no matter what the situation is. Then, if the attacker's best action is to attack, from the defender's point of view, the most profitable action is 1, and this choice should be made no matter what the circumstance is. Thus, the Nash equilibrium of this game is $da=11$.

\subsection{Probabilistic Model}
Different from the deterministic model, we now consider a situation closer to the reality, that the defender does not have to be fully deterministic with only auditing after sending the signal. Sometimes, out of some strategic considerations, the defender will not audit after sending the signal, or audit unexpectedly without sending a signal. In this case, we assume that there is a probability between the defender's signaling and auditing behavior, leading to the probabilistic model.

In the probabilistic model, we assume that if the defender sends a signal, the auditing will be done with the probability of $\tau$; otherwise, she audits with the probability of $\delta$, where $\tau>\delta$, as it is natural to be more inclined to audit when sending a signal.

Similar to the deterministic model, the probabilistic model also produces four possible states of the game between the defender and the attacker: $da=(00,01,10,11)$. The payoffs of four states are calculated as follows: i) for $da=00$, as the defender chooses not to send the signal and the attacker doesn't attack, the defender emerges an audit cost of $c$ with the probability of $\delta$ while the attacker acquires nothing; ii) for $da=01$, the defender chooses not to send the signal but the attacker attacks, the defender's loss consists of three parts: the audit cost, denoted as $c$, with the probability of $\delta$, the loss of being attacked without audited, denoted as $t_d$, with the probability of $1-\delta$, and the loss of being attacked but audited sooner, denoted as $t_m$, with the probability of $\delta$. Attacker gains income $r_a$ minus punishment $s_a$ from the audit with the probability of $\delta$; iii) for $da=10$, the defender sends a signal and audits with the probability of $\tau$ but the attacker quits, so the defender spends $c$ as the cost of auditing with the probability of $\tau$ and the attacker acquires nothing; iv) for $da=11$, denoting that the defender sends a signal and the attacker attacks. The defender's loss consists of three parts: the audit cost, denoted as $c$, with the probability of $\tau$, the loss of being attacked without audited, denoted as $t_d$, with a probability of $1-\tau$, the loss of being attacked but audited sooner, denoted as $t_m$, with the probability of $\tau$. While the attacker gains income $r_a$ minus punishment $s_a$ from the audit with the probability of $\tau$. Subsequently, the payoff vector of the defender is $\tilde{\mathbf{U}}^{\eta}_D=(-\delta c,-\delta c-(\delta t_m+(1-\delta)t_d),-\tau c,-\tau c-(\tau t_m+(1-\tau)t_d))$ and that for attacker is $\tilde{\mathbf{U}}^{\eta}_A=(0,r_a-\delta s_a,0, r_a-\tau s_a)$. For the sake of notation simplicity, we omit $\eta$ in the following expressions, using $\mathbf{U}_A$, $\mathbf{U}_D$, $\tilde{\mathbf{U}}_A$ and $\tilde{\mathbf{U}}_D$ instead of $\mathbf{U}^{\eta}_A$, $\mathbf{U}^{\eta}_D$, $\tilde{\mathbf{U}}^{\eta}_A$ and $\tilde{\mathbf{U}}^{\eta}_D$, respectively.

Other related restrictions are similar to those in the above subsection, but with an additional restriction $\tau>\delta$. Since the attacker acts secondly, he will choose 1 to make the largest profit. For the defender, we can also conclude that the benefit of choosing 1 is always greater. So the Nash equilibrium in the probabilistic model is still $da=11$.

\section{Utility Control of the Attacker using the Zero-Determinant Strategy}\label{Section4}

According to the analysis of Section \ref{Modeling}, we can see that there is a sequential Nash equilibrium in the game between the defender and the attacker, where the attacker’s optimal strategy is to attack because the attack always brings him positive benefits, while the defender’s optimal strategy is to send a signal and audit (with a higher probability in the probabilistic model) since auditing can effectively reduce the loss in both the deterministic model and the probabilistic model. In the long run, the defender consumes a lot of resources to send signals and conduct audits to play against potential attackers. However, considering that the defender's resource budget is generally limited, it is impossible to audit all requests including requests from non-attackers without restrictions. To solve this challenge, it becomes necessary to figure out an efficient strategy to audit requests, which can bring several benefits as follows. Firstly, this can effectively improve the audit efficiency and ensure the security of database information. Secondly, defender can also reduce the costs of signaling and auditing by sending signals strategically. In addition, reducing the number of signal prompts can improve the user experience for normal users.

In this section, we resort to the zero-determinant (ZD) strategy for achieving the above goals. Previous studies have proved that the ZD strategy ensures a linear relationship between the incomes of two players in the iterative game by setting an appropriate mixed strategy for one player, and even unilaterally set the opponent's expected income. This suggests us to propose a strategy to help the defender control the attacker’s utility and prevent the database from excessive damages. Nonetheless, the classic ZD strategy studies the simultaneous game between two parties without knowing each other's actions. Therefore, we need to expand the ZD strategy to our sequential games.

As mentioned in \cite{press2012iterated}, a long-memory player has no priority against a short-memory player in an iterated game. Therefore, we assume that the defender has only one round of memory. The defender's mixed strategy in a round is the conditional probability of choosing the strategy 0 based on all possible states of the previous round. As for the attacker, he has only one round of memory as well. His mixed strategy in a round is the conditional probability of choosing the strategy 0 based on all possible states of the previous round.

\begin{definition}(The defender's mixed strategy $\mathbf{p}$).
The mixed strategy of defender is denoted as $\mathbf{p}=(p_1,p_2,p_3,p_4)$, with each element being the probability of the defender to choose 0 when the outcome state of the previous round is $da=(00,01,10,11)$.
\end{definition}
Thus, $1-p_i$ $(i\in\{1,2,3,4\})$ denotes the probability of the defender to choose 1 when the outcome state of the previous round is $da=(00,01,10,11)$.
\begin{definition}(The attacker's mixed strategy $\mathbf{q}$).
The mixed strategy of the attacker is denoted as $\mathbf{q}=(q_1,q_2)$, with each element being the probability of the attacker to choose 0 when the defender's action in the current round is $d=(0,1)$.
\end{definition}
Respectively, $1-q_1$ and $1-q_2$ denote the probability of the attacker to choose 1 when the defender's action in this round is $d=(0,1)$.

Based on the above definitions, $\mathbf{p}$ and $\mathbf{q}$ can compose a Markov matrix denoting the state transition between two consecutive rounds, which can be expressed as:
\begin{equation*}\label{eq:markov_matrix}
\mathbf{M}=
\left[
  \begin{array}{cccc}
    p_1q_1 & p_1(1-q_1) & (1-p_1)q_2 & (1-p_1)(1-q_2)\\  
    p_2q_1 & p_2(1-q_1) & (1-p_2)q_2 & (1-p_2)(1-q_2)\\  
    p_3q_1 & p_3(1-q_1) & (1-p_3)q_2 & (1-p_3)(1-q_2)\\  
    p_4q_1 & p_4(1-q_1) & (1-p_4)q_2 & (1-p_4)(1-q_2)\\  
  \end{array}
\right].
\end{equation*}
Each element in $\mathbf{M}$ is the transition probability from the state in the last round to that in the current round. Taking the first row of $\mathbf{M}$ as an example, four elements denote the transition probabilities from state $da=00$ at the last round to the four possible states $da=00,01,10,11$ in the current round. The other three rows, similarly, correspond to the states $da=01,10,11$ in the last round.

We can easily calculate that $\mathbf{M}'\equiv \mathbf{M}-\mathbf{I}$ is singular with the determinant value of zero. Besides, the stationary vector of $\mathbf{M}$, denoted as $\mathbf{v}$, satisfies $\mathbf{v}^T\mathbf{M}=\mathbf{v}^T$ which equals $\mathbf{v}^T\mathbf{M}'=0$. Applying Cramer's rule on matrix $\mathbf{M}'$, we can get:
\begin{equation*}\label{eq:Cramer}
Adj(\mathbf{M}')\mathbf{M}'=\mathrm{det}(\mathbf{M}')\mathbf{I}=0,
\end{equation*}
where $Adj(\mathbf{M}')$ denotes the adjugate matrix of $\mathbf{M}'$. Thus, we can conclude that every row of $Adj(\mathbf{M}')$ is proportional to $\mathbf{v}$. The determinant of $\mathbf{M}'$ is unchanged if we add the first column of $\mathbf{M}'$ into the second and third columns. Thus, we can calculate the dot product of an arbitrary four-element vector $\mathbf{f}=(f_1,f_2,f_3,f_4)$ and the stationary vector $\mathbf{v}$ as follows:
\begin{align*}
\mathbf{v}\cdot \mathbf{f} \equiv D(\mathbf{p},\mathbf{q},\mathbf{f}) 
\end{align*}
\begin{align}\label{eq:vtimesf}
= \mathrm{det}
\left[
  \begin{array}{cccc}
    p_1q_1-1 & p_1-1 & (1-p_1)q_2+p_1q_1-1 & f_1\\  
    p_2q_1 & p_2-1 & (1-p_2)q_2+p_2q_1 & f_2\\  
    p_3q_1 & p_3 & (1-p_3)q_2+p_3q_1-1 & f_3\\  
    p_4q_1 & p_4 & (1-p_4)q_2+p_4q_1 & f_4\\  
  \end{array}
\right],
\end{align}
where the second column is under the control of the defender. Combining payoff vectors of the defender and the attacker, their respective utilities in the stationary state are:
\begin{equation*}\label{eq:u_a}
u_a=\frac{\mathbf{v}\cdot\mathbf{U}_A}{\mathbf{v}\cdot\mathbf{1}}=\frac{D(\mathbf{p},\mathbf{q},\mathbf{U}_A)}{D(\mathbf{p},\mathbf{q},\mathbf{1})},
\end{equation*}
\begin{equation*}\label{eq:u_d}
u_d=\frac{\mathbf{v}\cdot\mathbf{U}_D}{\mathbf{v}\cdot\mathbf{1}}=\frac{D(\mathbf{p},\mathbf{q},\mathbf{U}_D)}{D(\mathbf{p},\mathbf{q},\mathbf{1})}.
\end{equation*}
The above equations show that the utility of the attacker and that of the defender depend linearly on their corresponding payoff vectors. Thus, their linear combination of utilities will be calculated as:
\begin{equation}\label{eq:zd_equation}
\alpha u_a+\beta u_d+\gamma=\frac{D(\mathbf{p},\mathbf{q},\alpha \mathbf{U}_A+\beta \mathbf{U}_D+\gamma\mathbf{1})}{D(\mathbf{p},\mathbf{q},\mathbf{1})},
\end{equation}
with $\alpha,\beta,\gamma$ being constant parameters. It brings us many good attributes, allowing the defender to have a chance to make the determinant $D(\mathbf{p},\mathbf{q},\alpha \mathbf{U}_A+\beta \mathbf{U}_D+\gamma\mathbf{1})$ vanish. In fact, when the defender chooses a strategy that satisfies $\hat{\mathbf{p}}=\alpha \mathbf{U}_A+\beta \mathbf{U}_D+\gamma\mathbf{1}$, where $\hat{\mathbf{p}}$ denotes the second column of $D(\mathbf{p},\mathbf{q},\mathbf{f})$, the second column and the forth column of $D(\mathbf{p},\mathbf{q},\alpha \mathbf{U}_A+\beta \mathbf{U}_D+\gamma\mathbf{1})$ can be the same, then (\ref{eq:zd_equation}) changes to:
\begin{equation}\label{eq:zd_zero}
\alpha u_a+\beta u_d+\gamma=0.
\end{equation}
Thus, a linear relationship between $u_a$ and $u_d$ is enforced. The ZD strategy, however, is not feasible in all cases, which depends on whether the range of $\mathbf{p}$ is [0,1].

We can deploy the ZD strategy in the deterministic model and the probabilistic model, which provides the defender with a powerful approach to unilaterally control the attacker's utility.

\subsection{Deterministic Model}
In this part, we start with the basic deterministic model to find a strategy for the defender to control the attacker's utility. Generally, we analyze the relationship between the defender's strategy and the attacker's utility to get an appropriate strategy, and then find the most efficient variable to control the attacker's income, where the maximum and minimum utility of the attacker are analyzed as well to help the defender to assess potential risks.

From (\ref{eq:zd_zero}), we can see that the defender only needs to play a fixed strategy satisfying $\hat{\mathbf{p}}=\alpha\mathbf{U}_A+\gamma\mathbf{1}$ (setting $\beta=0$) to set the attacker’s utility. In this case, we can solve the below equation group:
\begin{equation}\label{eq:q1q2q3q4}
\begin{cases}   
    p_1-1=\gamma,\\  
    p_2-1=\alpha r_a+\gamma,\\  
    p_3=\gamma,\\  
    p_4=\alpha (r_a-s_a)+\gamma,
\end{cases}
\end{equation}
where $p_1$ and $p_4$ can be used to represent the remaining variables to get the expression of $u_a$:
\begin{equation}\label{eq:u_a_analysis}
u_a=-\frac{\gamma}{\alpha}=\frac{1-p_1}{p_4+1-p_1}\cdot(r_a-s_a).
\end{equation}
This expression implies that if the defender adopts a strategy satisfying $\hat{\mathbf{p}}=\alpha\mathbf{U}_A+\gamma\mathbf{1}$, the utility of the attacker can be determined by the defender. Then, we can analyze the features of $u_a$. Firstly, the value range of $u_a$ is $[0,r_a-s_a]$; secondly, in (\ref{eq:u_a_analysis}), $p_1$ and $p_4$ are variables that are unilaterally controlled by the defender, so we need to further study the extent of their influences on $u_a$. By this means, we can reveal that which variable is more effective to safeguard the system security to the greatest extent. Therefore, we first take the partial derivative of $u_a$ with respect to $p_1$,
\begin{equation}\label{eq:u_a'p1}
\frac{\partial u_a}{\partial p_1}=\frac{-p_4}{(p_4+1-p_1)^2}\cdot(r_a-s_a),
\end{equation}
where the derivative function decreases monotonically in $p_1\in[0,1]$. Further, we have:
\begin{equation*}
\overline{u}_a=u_a|(p_1=0)=\frac{1}{p_4}\cdot(r_a-s_a),\\
\end{equation*}
\begin{equation*}
\underline{u}_a=u_a|(p_1=1)=0,\\
\end{equation*}
where $\overline{u}_a$ denotes the maximum value of $u_a$ and $\underline{u}_a$ denotes the minimum value of $u_a$. This shows that if the defender only changes the value of $p_1$ in the strategy, the attacker's utility will be a certain value within the range of $[0,\frac{r_a-s_a}{p_4}]$.

Similarly, we take the partial derivative of $p_4$,
\begin{equation*}\label{eq:u_a'p4}
\frac{\partial u_a}{\partial p_4}=\frac{p_1-1}{(p_4+1-p_1)^2}\cdot(r_a-s_a).
\end{equation*}
It can be seen that the derivative function decreases monotonically in $p_4\in[0,1]$ and we have:
\begin{equation*}
\overline{u}_a=u_a|(p_4=0)=(r_a-s_a),
\end{equation*}
\begin{equation*}
\underline{u}_a=u_a|(p_4=1)=\frac{1-p_1}{2-p_1}\cdot(r_a-s_a).
\end{equation*}
This shows that if the defender only changes the value of $p_4$ in the strategy, the attacker's utility could be a certain value within the range of $[\frac{1-p_1}{2-p_1}\cdot(r_a-s_a),r_a-s_a]$.

To control the attacker’s utility more efficiently, we study which variable is more effective. In other words, when the increments of $p_1$ and $p_4$ are the same, which one of them causes a larger loss of the attacker's utility. Comparing the partial derivatives of two variables, we have:

\begin{equation*}\label{eq:comparingpartialderi1}
\frac{\partial u_a}{\partial p_1}-\frac{\partial u_a}{\partial p_4}=\frac{1-p_1-p_4}{(p_4+1-p_1)^2}\cdot(r_a-s_a).
\end{equation*}
It is clear that, when $1-p_1-p_4>0$, the partial derivative of $p_1$ is greater than that of $p_4$. Since they are all negative, it is more effective for the defender to control attacker's utility by changing $p_4$. While when $1-p_1-p_4<0$, the partial derivative of $p_4$ is greater than that of $p_1$. At this time, it is more effective for the defender to control attacker's utility by changing $p_1$.

Besides, $p_1$ and $p_4$ also have impacts on the value range of $u_a$. Regarding $p_1$ as the only variable, $u_a\in [0,\frac{r_a-s_a}{p_4}]$, with the range size of $\frac{r_a-s_a}{p_4}$. Regarding $p_4$ as the only variable, $u_a\in[\frac{1-p_1}{2-p_1}\cdot(r_a-s_a),r_a-s_a]$, with the range size of $\frac{r_a-s_a}{2-p_1}$. The above two sizes of range present the relationship of $\frac{r_a-s_a}{2-p_1}\le \frac{r_a-s_a}{p_4}$, since $2-p_1$ is in the range of $[1,2]$ and $p_4$ is in $[0,1]$, which means $p_1$ has a greater impact on the control range of $u_a$. Comparing the lower bounds of the above ranges, we have  $0\le\frac{1-p_1}{2-p_1}\cdot(r_a-s_a)$, while for the upper bounds, we have $\frac{r_a-s_a}{p_4}\le r_a-s_a$. Thus, if the defender tries to control $u_a$ at a low level, it is more effective to change $p_1$. 

According to the analysis above, we can conclude that when $p_1<1-p_4$, $p_1$ has a greater impact on the value of $u_a$; when $p_1>1-p_4$, $p_4$ has a greater impact on the value of $u_a$. In order to deploy defense strategies more effectively, the defender should pay attention to the relationship between the $p_1+p_4$ and 1. And if the defender can only change one variable, changing $p_1$ can be more conducive to limit the attacker's utility. 

\subsection{Probabilistic Model}

Similarly, we can analyze the probabilistic model. It should be noted that the two newly added variables $\tau$ and $\delta$ in the probabilistic model are unilaterally controlled by the defender, because they are used to determine the probability of auditing after signaling. Although $\tau$ and $\delta$ are different in definition from the strategy vector $\mathbf{p}$, their property of being controlled by the defender implies that they are also worthy of being studied. Solving the equation group like (\ref{eq:q1q2q3q4}), the expression of $\tilde{u}_a$ in the probabilistic model becomes:
\begin{equation}\label{eq:p_tilde_u_a}
\tilde{u}_a=-\frac{\gamma}{\alpha}=\frac{1-p_1}{p_4+1-p_1}\cdot(r_a-\tau s_a).
\end{equation}

Clearly, the value range of $\tilde{u}_a$ is $[0,r_a-\tau s_a]$. Further, in order to allow the defender to control the attacker’s utility $\tilde{u}_a$ more efficiently, we study the influence of the four variables $p1,p4,\tau$ and $\delta$ controlled by the defender on $\tilde{u}_a$ from a mathematical perspective. Notice that only $p1,p4$ and $\tau$ appear in (\ref{eq:p_tilde_u_a}), so we ignore the effect of $\delta$ and take the partial derivative of $\tilde{u}_a$ with respect to $p_1$ firstly: 
\begin{equation}\label{eq:tilde_u_a'p1}
\frac{\partial \tilde{u}_a}{\partial p_1}=\frac{-p_4}{(p_4+1-p_1)^2}\cdot(r_a-\tau s_a).
\end{equation}
From (\ref{eq:tilde_u_a'p1}), the derivative function decreases monotonically in $p_1\in[0,1]$. If we regard $p_1$ as the only variable, then we have:
\begin{equation*}
\overline{\tilde{u}}_a=\tilde{u}_a|(p_1=0)=\frac{1}{p_4}\cdot(r_a-\tau s_a),\\
\end{equation*}
\begin{equation*}
\underline{\tilde{u}}_a=\tilde{u}_a|(p_1=1)=0,\\
\end{equation*}
where $\overline{\tilde{u}}_a$ denotes the maximum value of $\tilde{u}_a$, while $\underline{\tilde{u}}_a$ denotes the minimum value of $\tilde{u}_a$. It can be seen that if the defender only changes the value of $p_1$ in the strategy, the attacker's utility will be a certain value within the range of $[0,\frac{r_a-\tau s_a}{p_4}]$.

Similarly, taking the derivative of $p_4$, we have:
\begin{equation*}\label{eq:tilde_u_a'p4}
\frac{\partial \tilde{u}_a}{\partial p_4}=\frac{p_1-1}{(p_4+1-p_1)^2}\cdot(r_a-\tau s_a),
\end{equation*}
where the derivative function decreases monotonically in $p_4\in[0,1]$. Regarding $p_4$ as a variable, we have:
\begin{equation*}
\overline{\tilde{u}}_a=\tilde{u}_a|(p_4=0)=r_a-\tau s_a,
\end{equation*}
\begin{equation*}
\underline{\tilde{u}}_a=\tilde{u}_a|(p_4=1)=\frac{1-p_1}{2-p_1}\cdot(r_a-\tau s_a),
\end{equation*}
which shows that if the defender only changes the value of $p_4$ in the strategy, the attacker's utility will be a certain value within the range of $[\frac{1-p_1}{2-p_1}\cdot(r_a-\tau s_a),r_a-\tau s_a]$.

In the probabilistic model, the effect of $p_1$ and $p_4$ are similar to that in the deterministic model. By comparing the partial derivatives of two variables:
\begin{equation*}\label{eq:comparingpartialderi2}
\frac{\partial u_a}{\partial p_1}-\frac{\partial u_a}{\partial p_4}=\frac{1-p_1-p_4}{(p_4+1-p_1)^2}\cdot(r_a-\tau s_a),
\end{equation*}
we can draw the same conclusion with that in the deterministic model: when $p_1<1-p_4$, $p_1$ has a greater impact on the control of the value of $\tilde{u}_a$; when $p_1>1-p_4$, $p_4$ is more effective to control $\tilde{u}_a$. Thus, $p_1$ has a greater impact on the value range of $\tilde{u}_a$ as well as on controlling $\tilde{u}_a$ at a low level.

Meanwhile, $p_1$ and $p_4$ have impacts on the value range of $\tilde{u}_a$. Regard $p_1$ as the only variable, $\tilde{u}_a\in [0,\frac{r_a-\tau s_a}{p_4}]$, with the range of $\frac{r_a-\tau s_a}{p_4}$. Regard $p_4$ as the only variable, $\tilde{u}_a\in[\frac{1-p_1}{2-p_1}\cdot(r_a-\tau s_a),r_a-\tau s_a]$, with the range of $\frac{r_a-\tau s_a}{2-p_1}$. As for the size of range, we have $\frac{r_a-\tau s_a}{2-p_1}\le \frac{r_a-\tau s_a}{p_4}$ as $2-p_1\in[1,2]$ and $p_4\in[0,1]$. Comparing the lower bounds of the above ranges, we have $0\le\frac{1-p_1}{2-p_1}\cdot(r_a-\tau s_a)$, while for the upper bounds, we have $\frac{r_a-\tau s_a}{p_4}\le r_a-\tau s_a$, so $p_1$ has greater influence on controlling value of $\tilde{u}_a$.

In addition, the influence of $\tau$ on $\tilde{u}_a$ is different from that of $p_1$ and $p_4$, as the partial derivatives of $\tau$ is:
\begin{equation*}\label{eq:tilde_u_a'tau}
\frac{\partial \tilde{u}_a}{\partial \tau}=\frac{s_a(p_1-1)}{(p_4+1-p_1)},
\end{equation*}
which means the relationship between $\tau$ and $\tilde{u}_a$ is negative correlated since $p_1\le1$. If we only regard $\tau$ as a variable, we have:
\begin{equation*}
\overline{\tilde{u}}_a=\tilde{u}_a|(\tau=0)=\frac{r_a(1-p_1)}{p_4+1-p_1},
\end{equation*}
\begin{equation*}
\underline{\tilde{u}}_a=\tilde{u}_a|(\tau=1)=\frac{1-p_1}{p_4+1-p_1}\cdot(r_a-s_a),
\end{equation*}
where $\overline{\tilde{u}}_a$ denotes the maximum value of $\tilde{u}_a$, while $\underline{\tilde{u}}_a$ denotes the minimum value of $\tilde{u}_a$. This shows that if the defender only changes the value of $\tau$ in the strategy, the attacker's utility will be a certain value within the range of $[\frac{(1-p_1)}{p_4+1-p_1}\cdot(r_a-s_a),\frac{r_a(1-p_1)}{p_4+1-p_1}]$. Regard $\tau$ as the only variable, $\tilde{u}_a\in[\frac{(1-p_1)}{p_4+1-p_1}\cdot(r_a-s_a),\frac{r_a(1-p_1)}{p_4+1-p_1}]$, whose range is $\frac{s_a(1-p_1)}{p_4+1-p_1}$.

\section{Maximizing the Utility Difference using the Zero-Determinant Strategy}\label{Section5}

The ZD strategy demonstrates powerful control over the attacker's utility as mentioned in the previous section. Although controlling the attacker's utility sometimes leads to excellent performance, simply controlling the attacker's utility at a lower level may result in huge budget expenditures. Therefore, when necessary, we hope to design a strategy that considers both the utility of the defender and that of the attacker. Different from the defender's utility, the utility difference between the defender and the attacker is a relative value, and the study of utility difference is helpful for the defender to flexibly deal with the strategies of different attackers. Because this repeated game is not a zero-sum game, if the defender has the highest utility, the attacker could be likely to get a high utility as well, which can bring more damage to the database. In this section, we use the ZD strategy to find the maximum utility difference in defender's point of view.

Our main idea is to propose a set of signal and audit strategies for the defender to make $\tilde{u}_d-\tilde{u}_a$ the largest. It should be noted that although we proposed two models before, i.e., the deterministic model and the probabilistic model, in this section, we use the probabilistic model as an example to explore the utility difference control. The reason is that compared with the deterministic model, the variables $\tau$ and $\delta$ in the probabilistic model expand the action space of the defender, which is more flexible and comprehensive. In addition, it is easy to get similar conclusions in the probabilistic model and the deterministic model, via eliminating the influence of $\tau$ and $\delta$ by setting $\tau=1$ and $\delta=0$.

According to (\ref{eq:zd_zero}), by setting $\alpha=-1$ and $\beta=1$, the utility difference between the defender and the attacker can be calculated as:
\begin{equation*}\label{eq:max_ud-ua}
\tilde{u}_d-\tilde{u}_a=-\gamma.
\end{equation*}
Hence, the basic issue of maximizing the utility difference can
be achieved by solving:
\begin{equation*}\label{eq:basicopt}
\begin{split}
&\max\, -\gamma,\\
&\,s.t.\quad 0\le p_i\le 1,\forall{i}\in\{1,2,3,4\},\\
\end{split}
\end{equation*}
which is equivalent to the following optimization problem with constraints:
\begin{equation*}\label{eq:min_gamma}
\begin{split}
&\min \gamma,\\
&\,s.t.
\begin{cases}
0\le p_i\le 1,\forall{i}\in\{1,2,3,4\},\\
\hat{\mathbf{p}}=\phi(\tilde{\mathbf{U}}_D- \tilde{\mathbf{U}}_A+\gamma\mathbf{1}),\\
\phi \neq 0.\\
\end{cases}
\end{split}
\end{equation*}
Among them, $\hat{\mathbf{p}}=(p_1-1,p_2-1,p_3,p_4)$ is the second column in (\ref{eq:vtimesf}), which can be unilaterally determined by the defender's strategy. We denote $\tilde{U}^k_A$ and $\tilde{U}^k_D$ as the $k$th element in $\tilde{\mathbf{U}}_A$ and $\tilde{\mathbf{U}}_D$, respectively. Then we can solve the above optimization problem by considering the following two cases:
\subsubsection{Case 1}$\phi>0$.
To meet the constraint $p_i\ge0$, we can get the lower bound of $\gamma$ as follows:
\begin{align*}
& {\gamma}_{min}=\max(\Lambda_k),\forall{k}\in\{1,2,3,4\},\\
& \Lambda_k=
\begin{cases}   
-\tilde{U}^k_D+\tilde{U}^k_A-\frac{1}{\phi}, &k=1,2,\\
-\tilde{U}^k_D+\tilde{U}^k_A, & k=3,4.\\
\end{cases}
\end{align*}
To meet the constraint $p_i\le1$, we can get the upper bound of $\gamma$ as follows:
\begin{align*}
& {\gamma}_{max}=\min(\Lambda_l),\forall{l}\in\{5,6,7,8\},\\
& \Lambda_l= \Lambda_{k+4}
\begin{cases}
-\tilde{U}^k_D+\tilde{U}^k_A, &k=1,2,\\
-\tilde{U}^k_D+\tilde{U}^k_A+\frac{1}{\phi}, & k=3,4.\\
\end{cases}
\end{align*}
Only if $\gamma_{min}\le\gamma_{max}$ can $\gamma$ has a feasible solution, which is equivalent to $\max(\Lambda_k)\le \min(\Lambda_l),\forall{k}\in\{1,2,3,4\},\forall{l}\in\{5,6,7,8\}$. If there exists $\phi>0$ satisfying the above constraint, we can obtain the minimum value of $\gamma$ as follow:
\begin{multline}\label{eq:case1_gammamin}
\gamma_{min}=\max\{-\tilde{U}^1_D+\tilde{U}^1_A-\frac{1}{\phi},-\tilde{U}^2_D+\tilde{U}^2_A-\frac{1}{\phi},\\
-\tilde{U}^3_D+\tilde{U}^3_A,-\tilde{U}^4_D+\tilde{U}^4_A\} \\
=\max\{\delta c-\frac{1}{\phi},\delta c+(\delta t_m+(1-\delta)t_d)+r_a-\delta s_a-\frac{1}{\phi},\\
\tau c,\tau c+(\tau t_m+(1-\tau)t_d)+r_a-\tau s_a\}.
\end{multline}

\subsubsection{Case 2}$\phi<0$.
Similarly, when considering that $p_i\ge0$, we have $\gamma_{min}=\max(\Lambda_l),\forall{l}\in\{5,6,7,8\}$; while when considering that $p_i\le1$, we have $\gamma_{max}=\min(\Lambda_k),\forall{k}\in\{1,2,3,4\}$. In addition, $\gamma$ is feasible only when $\gamma_{min}\le\gamma_{max}$, i.e., $\max(\Lambda_l)\le \min(\Lambda_k),\forall{k}\in\{1,2,3,4\},\forall{l}\in\{5,6,7,8\}$. Finally, we can get the following result:
\begin{multline}\label{eq:case2_gammamin}
\gamma_{min}=\max\{-\tilde{U}^1_D+\tilde{U}^1_A,-\tilde{U}^2_D+\tilde{U}^2_A,\\
-\tilde{U}^3_D+\tilde{U}^3_A+\frac{1}{\phi},-\tilde{U}^4_D+\tilde{U}^4_A+\frac{1}{\phi}\}\\
=\max\{\delta c,\delta c+(\delta t_m+(1-\delta)t_d)+r_a-\delta s_a,\\
\tau c+\frac{1}{\phi},\tau c+(\tau t_m+(1-\tau)t_d)+r_a-\tau s_a+\frac{1}{\phi}\}.
\end{multline}
In summary, by (\ref{eq:case1_gammamin}) and (\ref{eq:case2_gammamin}), the defender can unilaterally set the maximum value of $\tilde{u}_d-\tilde{u}_a$ with the ZD strategy $\mathbf{p}$ meeting $\hat{\mathbf{p}}=\phi(\tilde{\mathbf{U}}_D- \tilde{\mathbf{U}}_A+\gamma\mathbf{1})$, where each element of $\mathbf{p}$ can be calculated by:
\begin{align*}\label{eq:p_i_calculate}
p_i=
\begin{cases}   
\tilde{U}^i_D-\tilde{U}^i_A+\gamma_{min}+1, &i=1,2,\\
\tilde{U}^i_D-\tilde{U}^i_A+\gamma_{min}, & i=3,4.\\
\end{cases}
\end{align*}
\begin{remark}
For the deterministic model, we can have the optimization problem as:
\begin{equation*}\label{eq:min_gamma2}
\begin{split}
& \min \gamma,\\
& s.t.
\begin{cases}
0\le p_i\le 1,\forall{i}\in\{1,2,3,4\},\\
\hat{\mathbf{p}}=\phi(\mathbf{U}_D- \mathbf{U}_A+\gamma\mathbf{1}),\\
\phi \neq 0,\\
\end{cases}
\end{split}
\end{equation*}
which can be easily solved using the above conclusions. Specifically, we can derive the maximized utility difference as:
\begin{multline*}\label{eq:p_i_calculate2}
\gamma_{min}=
\begin{cases}   
\max\{-U^1_D+U^1_A-\frac{1}{\phi},-U^2_D+U^2_A-\frac{1}{\phi},\\
\qquad -U^3_D+U^3_A,-U^4_D+U^4_A\},\qquad \phi>0,\\
\max\{-U^1_D+U^1_A,-U^2_D+U^2_A,-U^3_D+U^3_A+\frac{1}{\phi},\\
\qquad -U^4_D+U^4_A+\frac{1}{\phi}\},\qquad \phi<0,\\
\end{cases}
\end{multline*}
and $\mathbf{p}$ is given by    
\begin{align*}
p_i=
\begin{cases}   
U^i_D-U^i_A+\gamma_{min}+1, &i=1,2,\\
U^i_D-U^i_A+\gamma_{min}, & i=3,4.\\
\end{cases}
\end{align*}
\end{remark}

\section{Experimental Evaluation}

In this section, we evaluate the performance of the ZD strategy for the defender based on simulation experiments. All experiments are implemented using Matlab R2020a on a laptop with 2.3 GHz Intel Core i5-8300H processor. Besides, for the common parameters of the deterministic model and probabilistic model, we set the following default values: the loss of non-auditing after being attacked $t_d=8$, the loss of auditing after being attacked $t_m=5$; the income of the successful attack $r_a=10$, the loss of the attack being audited $s_a=5$; the cost of auditing $c=2$. Each experiment is repeated 50 times to get the average results for statistical confidence. We also conduct multiple experiments with different parameter settings, but all the experimental results are similar or have the same statistical significance. Therefore, in order to  avoid redundancy, we omit them and report the most representative experimental results.

\subsection{Unilateral Control of the Attacker's Utility using the ZD Strategy}

\begin{figure}
  \centering
  \subfigure[]{
    \label{fig:subfig:3a}
    \includegraphics[scale=0.286]{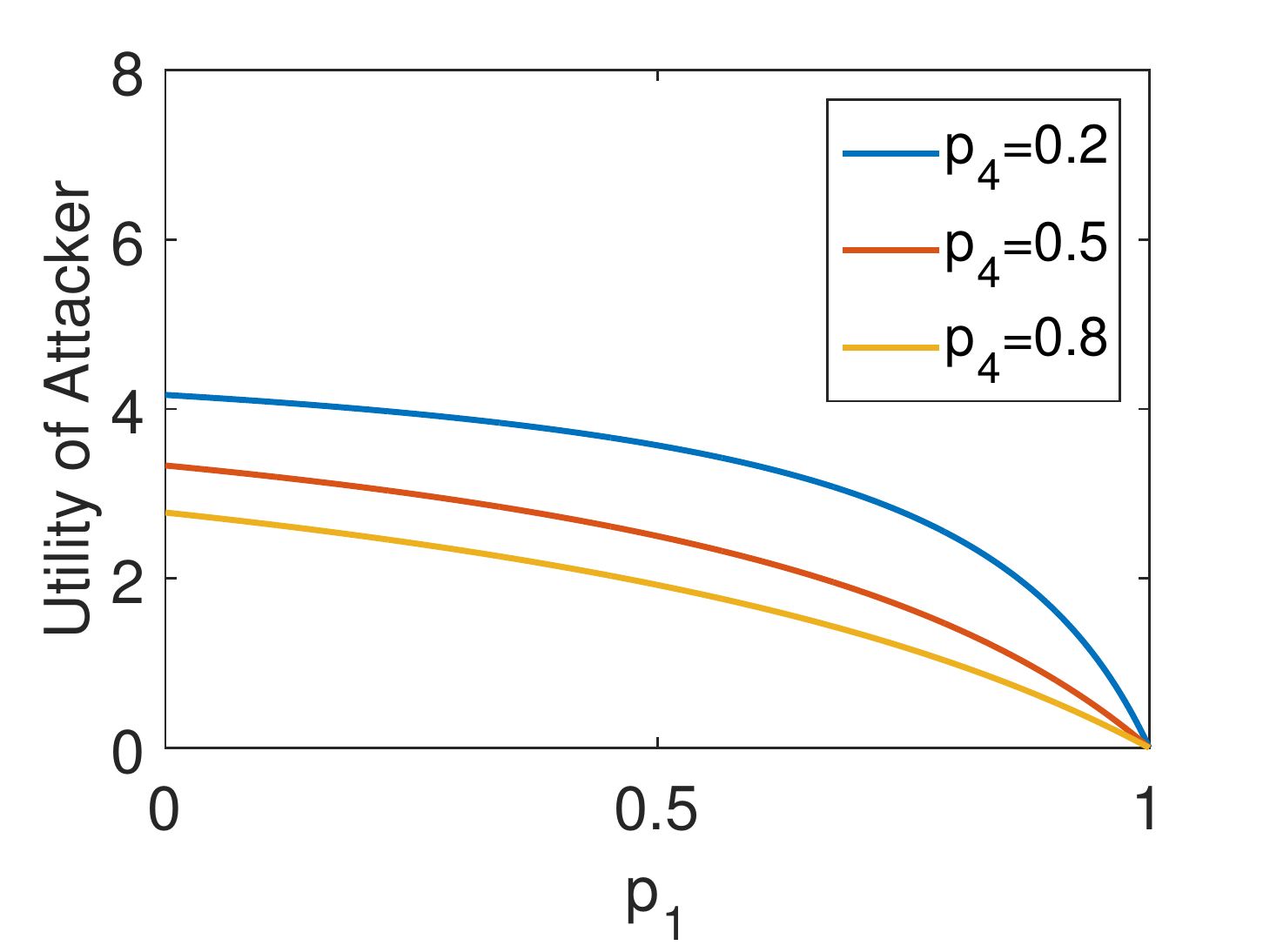}}
  \subfigure[]{
    \label{fig:subfig:3b}
    \includegraphics[scale=0.286]{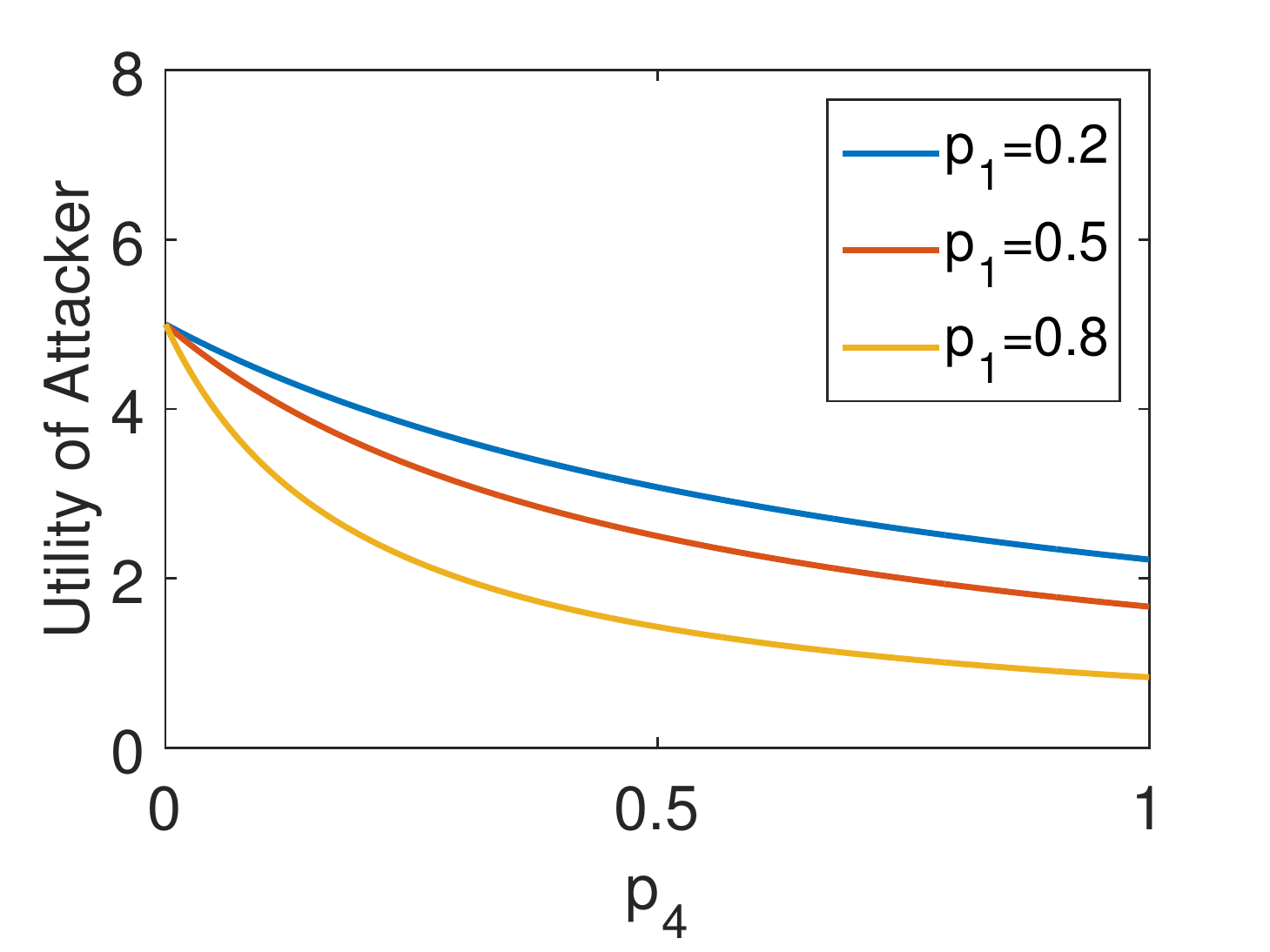}}
  \caption{The attacker's utility changes with different defender's strategy variables in the deterministic model.}
  \label{fig:deterministicp1p4}
\end{figure}

\begin{figure}
  \centering
  \subfigure[]{
    \label{fig:subfig:4a}
    \includegraphics[scale=0.286]{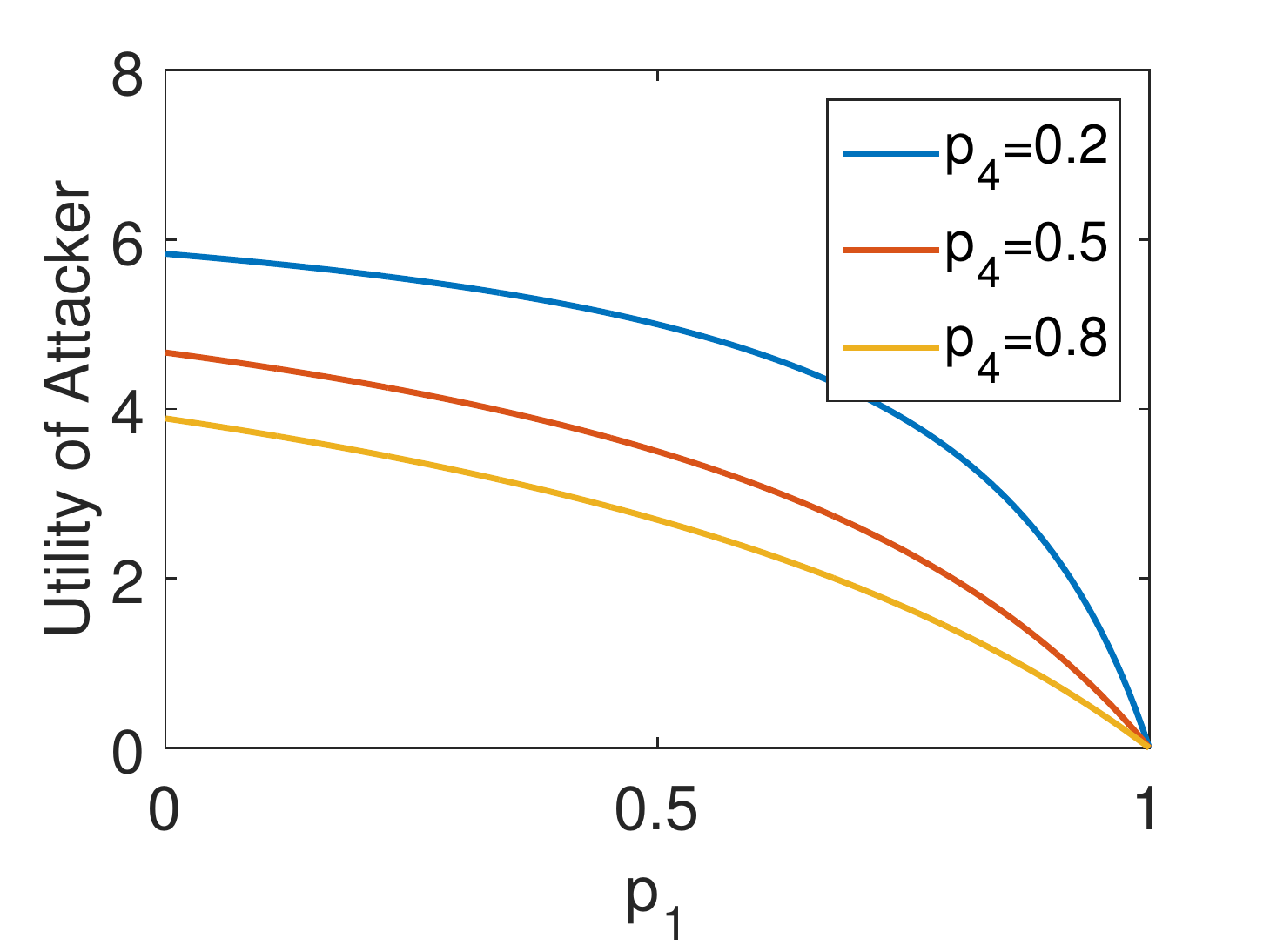}}
  \subfigure[]{
    \label{fig:subfig:4b}
    \includegraphics[scale=0.286]{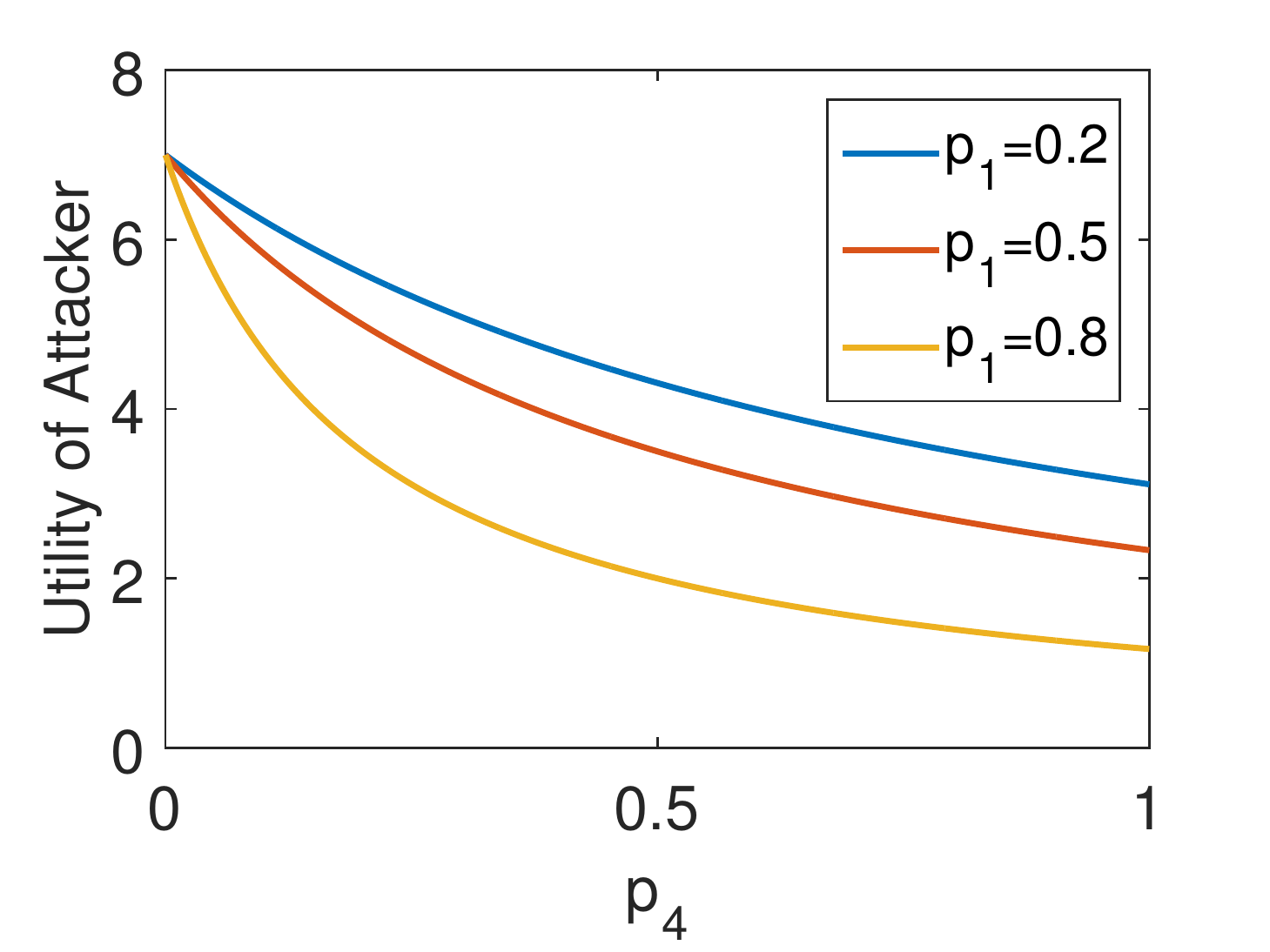}}
  \subfigure[]{
    \label{fig:subfig:4c}
    \includegraphics[scale=0.286]{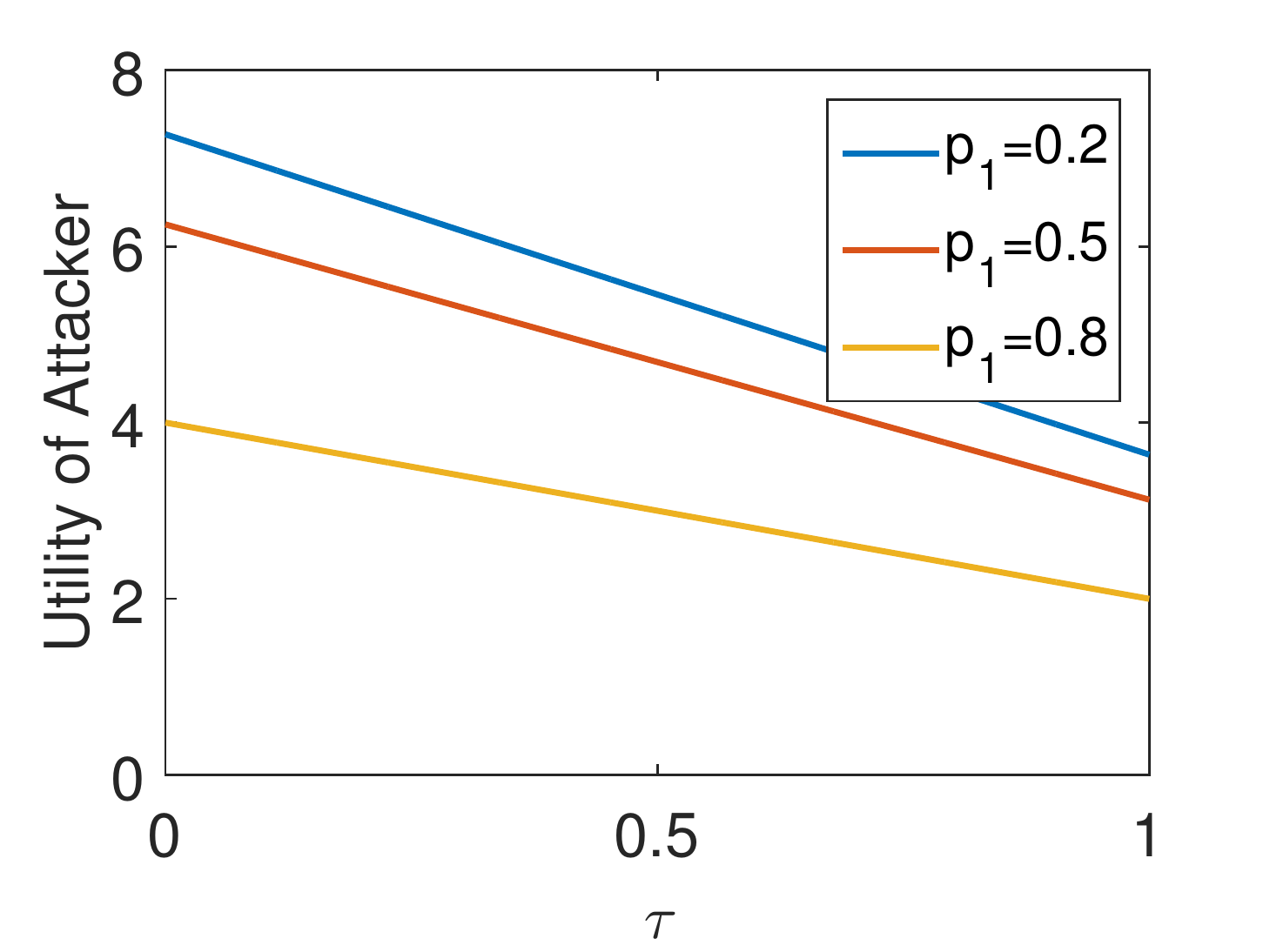}}
  \subfigure[]{
    \label{fig:subfig:4d}
    \includegraphics[scale=0.286]{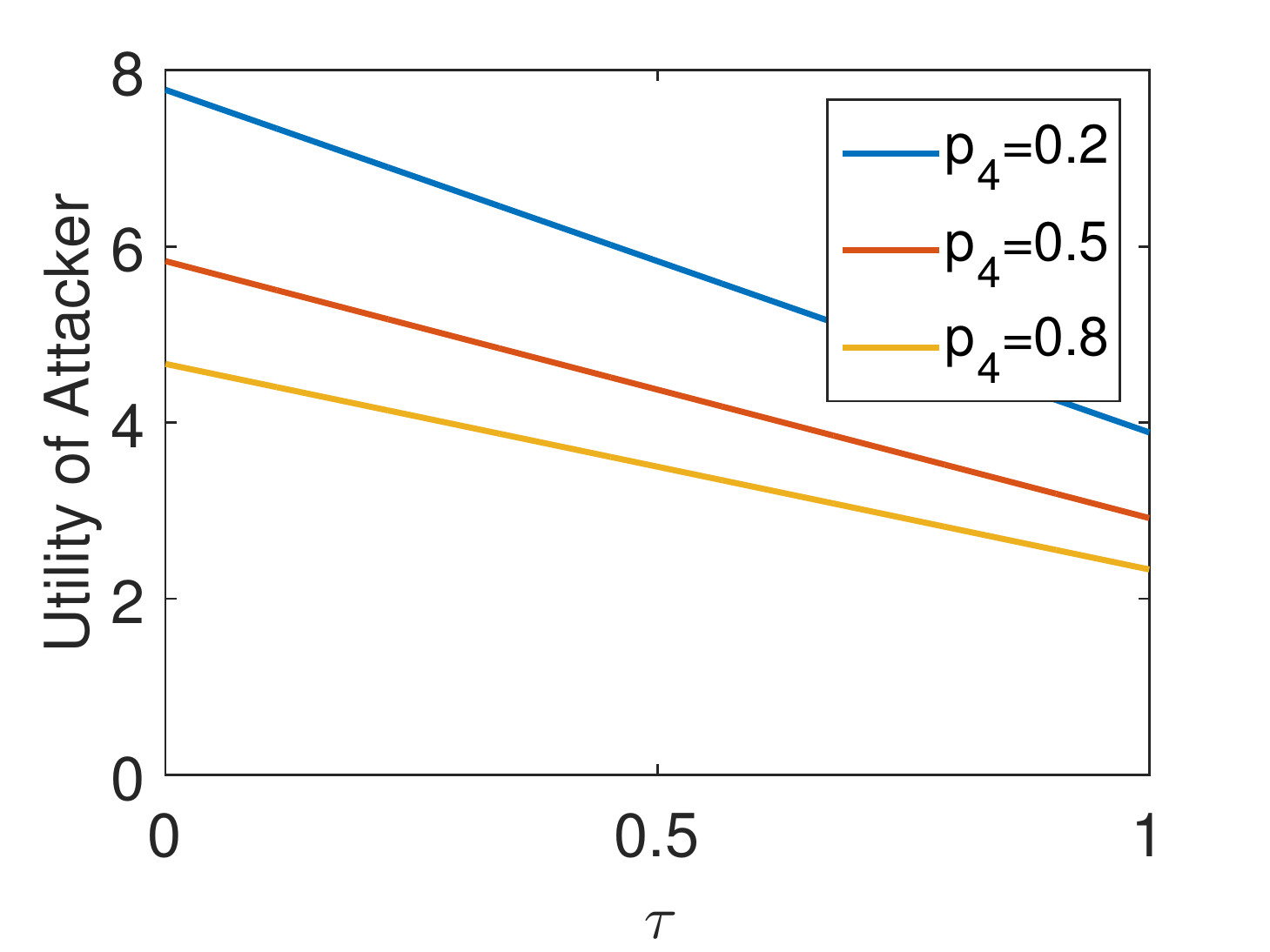}}
  \caption{The attacker's utility changes with different defender's strategy variables in the probabilistic model.}
  \label{fig:probabilisticp1p4tau}
\end{figure}

We deploy simulation experiments to verify the effectiveness of the defender using the ZD strategy to unilaterally control the attacker's utility, as well as demonstrate how the defender controls the attacker's utility based on $p_1$ and $p_4$. Fig. \ref{fig:deterministicp1p4} plots the attacker's utility changing with the defender's various strategy variables in the deterministic model. As mentioned in Section \ref{Section4}, $p_1$ and $u_a$ are negatively correlated. The changing rate increases as $p_1$ increases. And $p_4$, is also negatively correlated with $u_a$ while the rate of change decreases as $p_4$ increases. Fig. \ref{fig:probabilisticp1p4tau} presents that the probabilistic model has similar properties. It is worth noting that Fig. \ref{fig:probabilisticp1p4tau}(c)(d) shows that $\tau$ has a linear relationship with $\tilde{u}_a$, where the higher the $\tau$, the lower the attacker's utility. 

In addition, to verify the effectiveness of our scheme, we compare the defender's ZD strategy with other five classic strategies. We simulate the entire process of the defender and the attacker in the deterministic model, for 50 rounds, in which the defender uses the ZD, All-Zero (ALL0)\cite{hu2019quality}, All-One (ALL1)\cite{hu2019quality}, Random (Rand)\cite{hu2019quality}, Tit-For-Tat (TFT)\cite{nowak1993strategy}, and Win-Stay-Lose-Shift (WSLS)\cite{posch1997win} strategies. The attacker adopts ALL0, ALL1, Rand, TFT, and WSLS strategies. Specifically, ALL0 strategy is defined as: the defender always takes the action of not sending the signal no matter what the opponent does and the attacker always chooses to quit. ALL1 strategy means that the defender always sends the signal and the attacker always chooses to attack. With the Rand strategy, each player selects the action of 0 with the probability of 0.5. TFT strategy is defined as the player follows the choice of the opponent in the previous round. While WSLS strategy is defined as the player follows the choice if it won in the previous round, but changes to the other action otherwise.

\begin{figure}[h]
\centering
  \centering
  \subfigure[ZD]{
    \label{fig:subfig:5a}
    \includegraphics[scale=0.286]{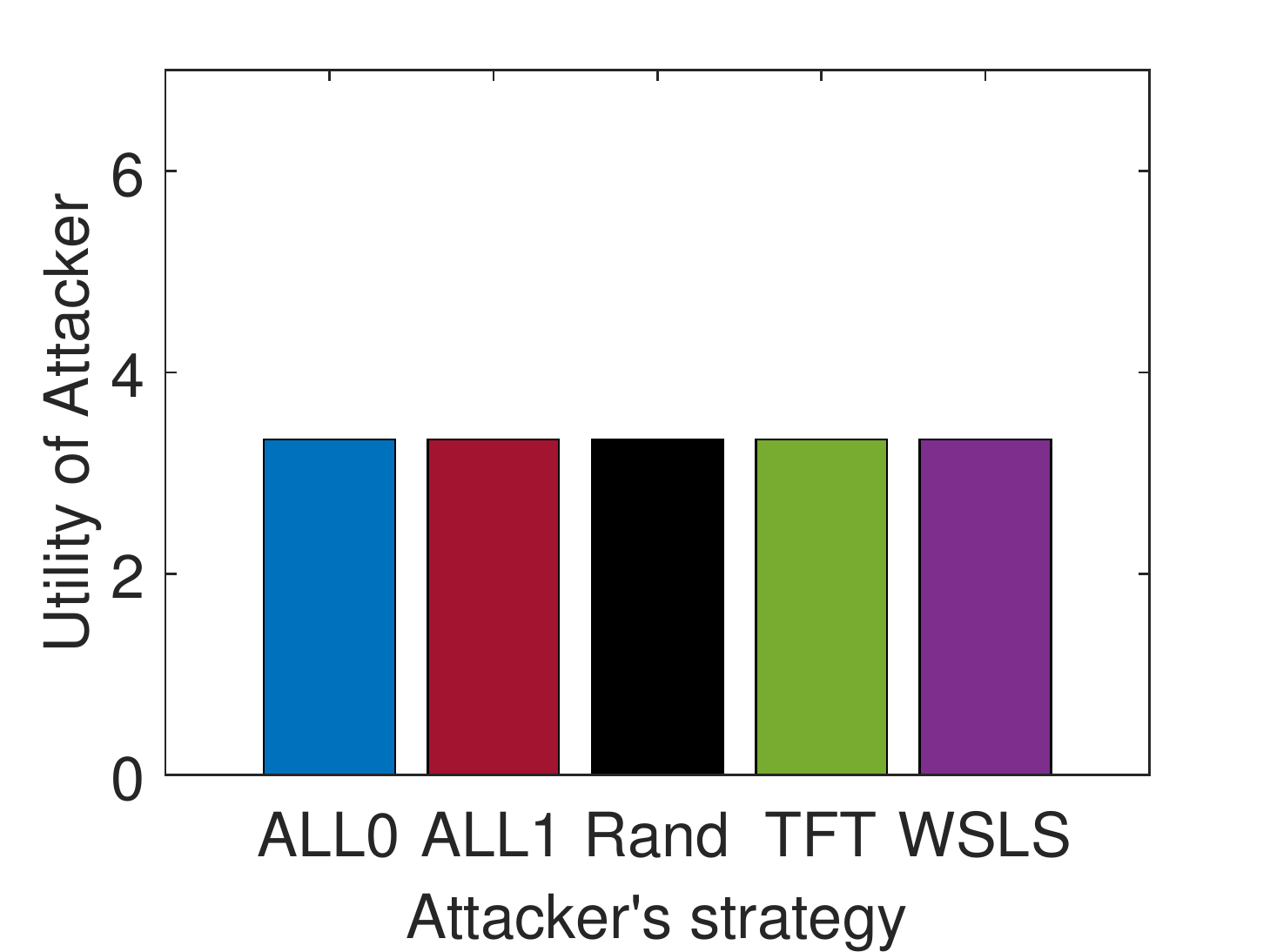}}
  \subfigure[ALL0]{
    \label{fig:subfig:5b}
    \includegraphics[scale=0.286]{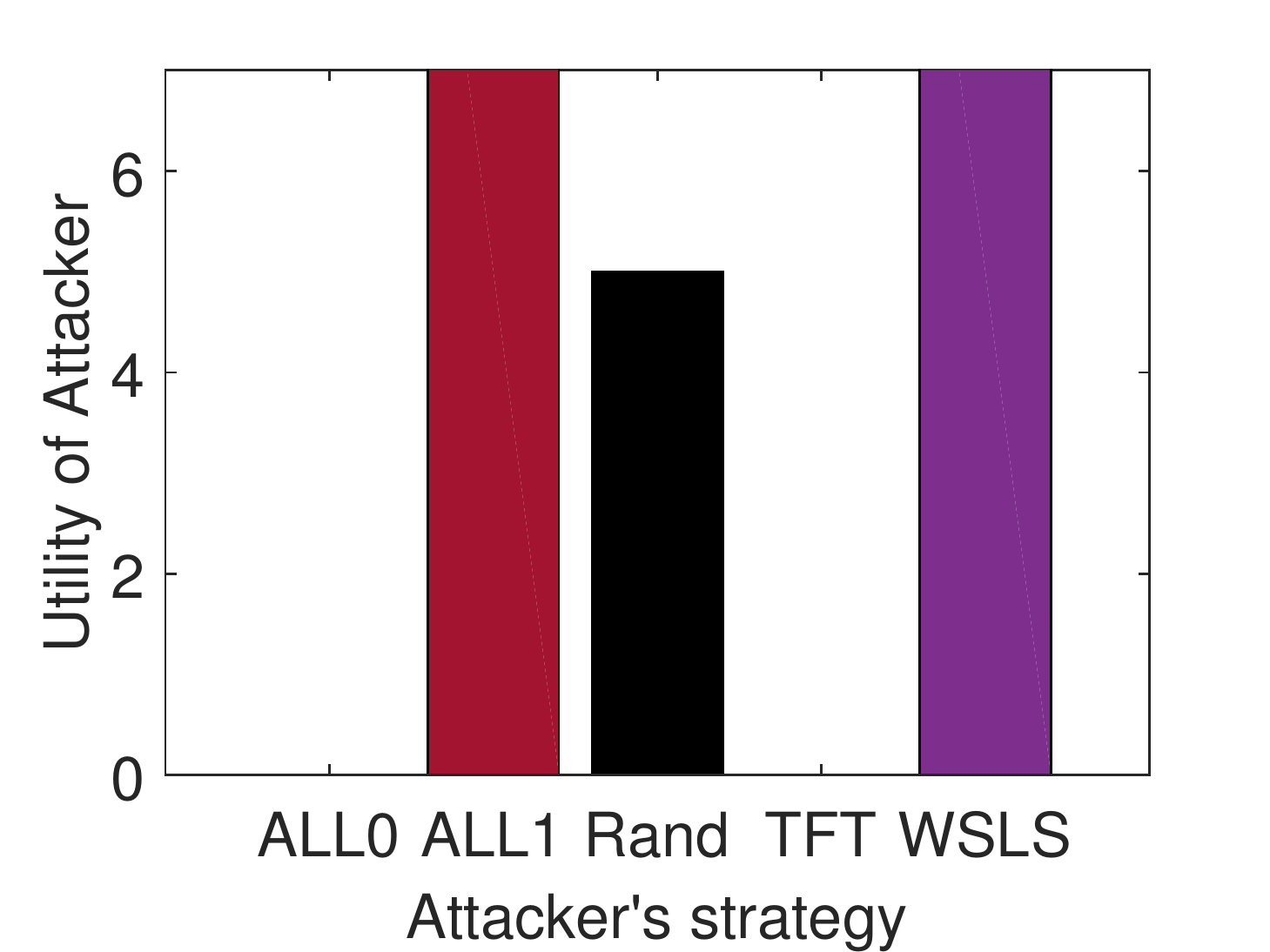}}
  \subfigure[ALL1]{
    \label{fig:subfig:5c}
    \includegraphics[scale=0.286]{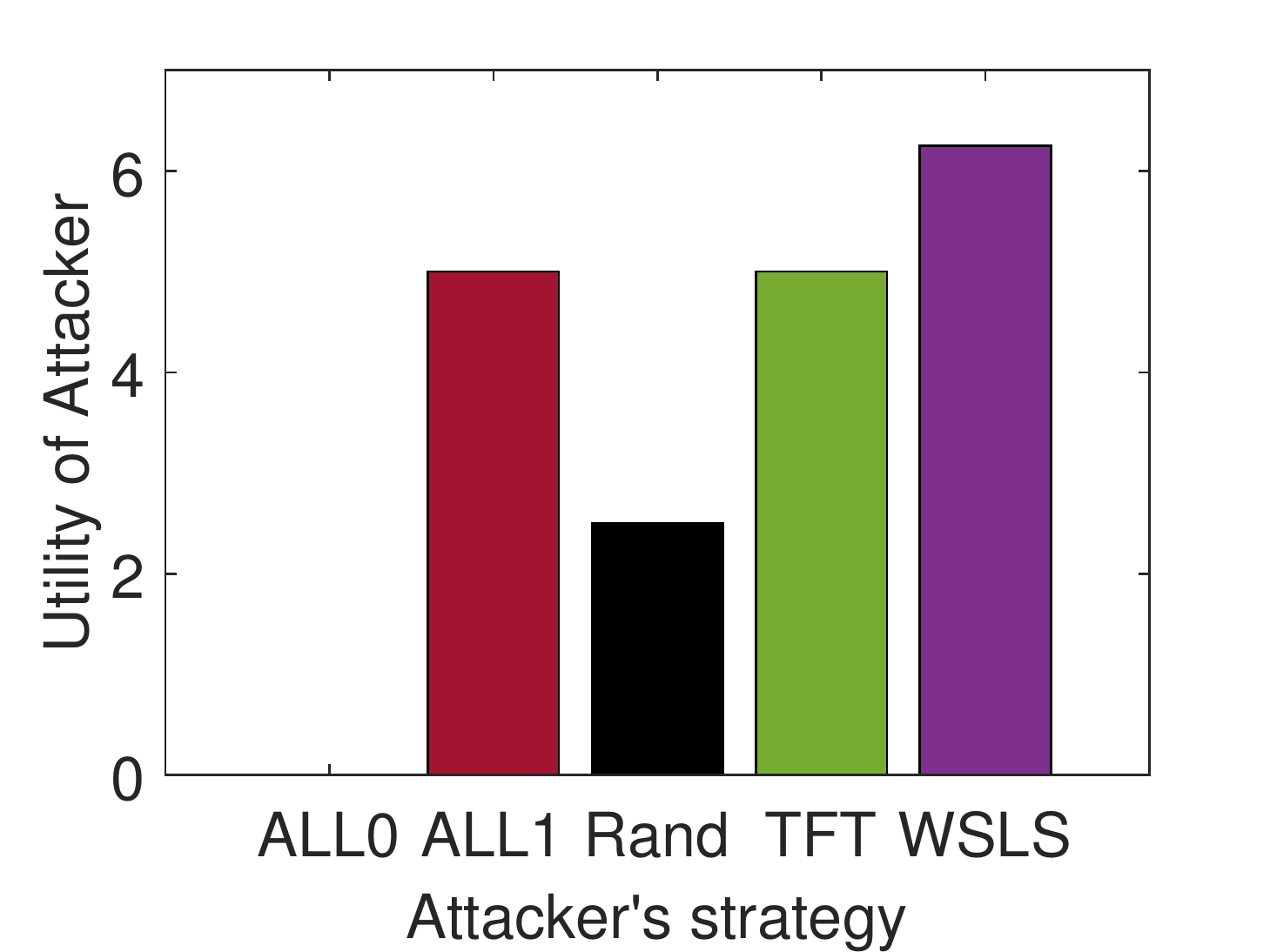}}
  \subfigure[Rand]{
    \label{fig:subfig:5d}
    \includegraphics[scale=0.286]{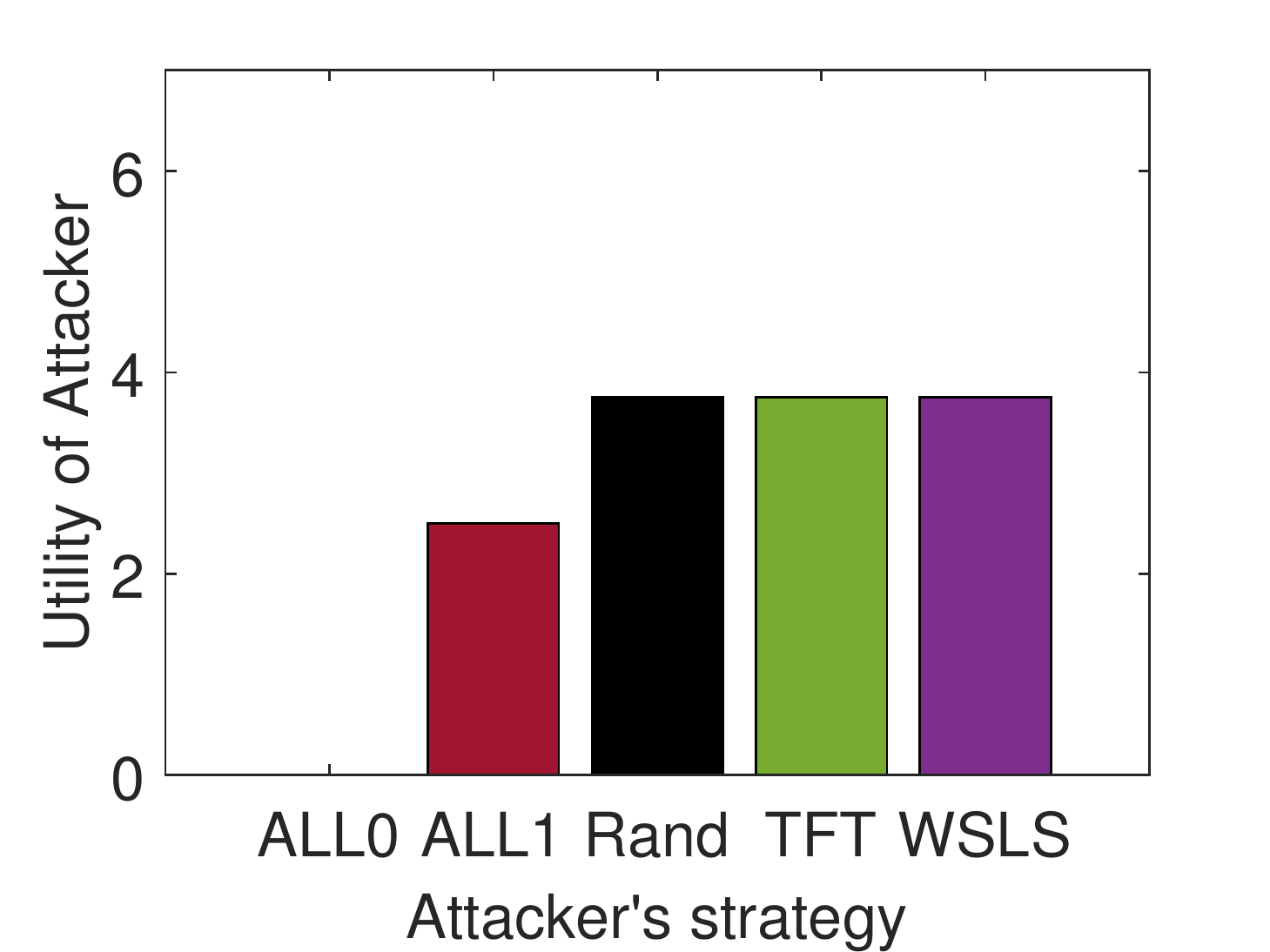}}
  \subfigure[TFT]{
    \label{fig:subfig:5e}
    \includegraphics[scale=0.286]{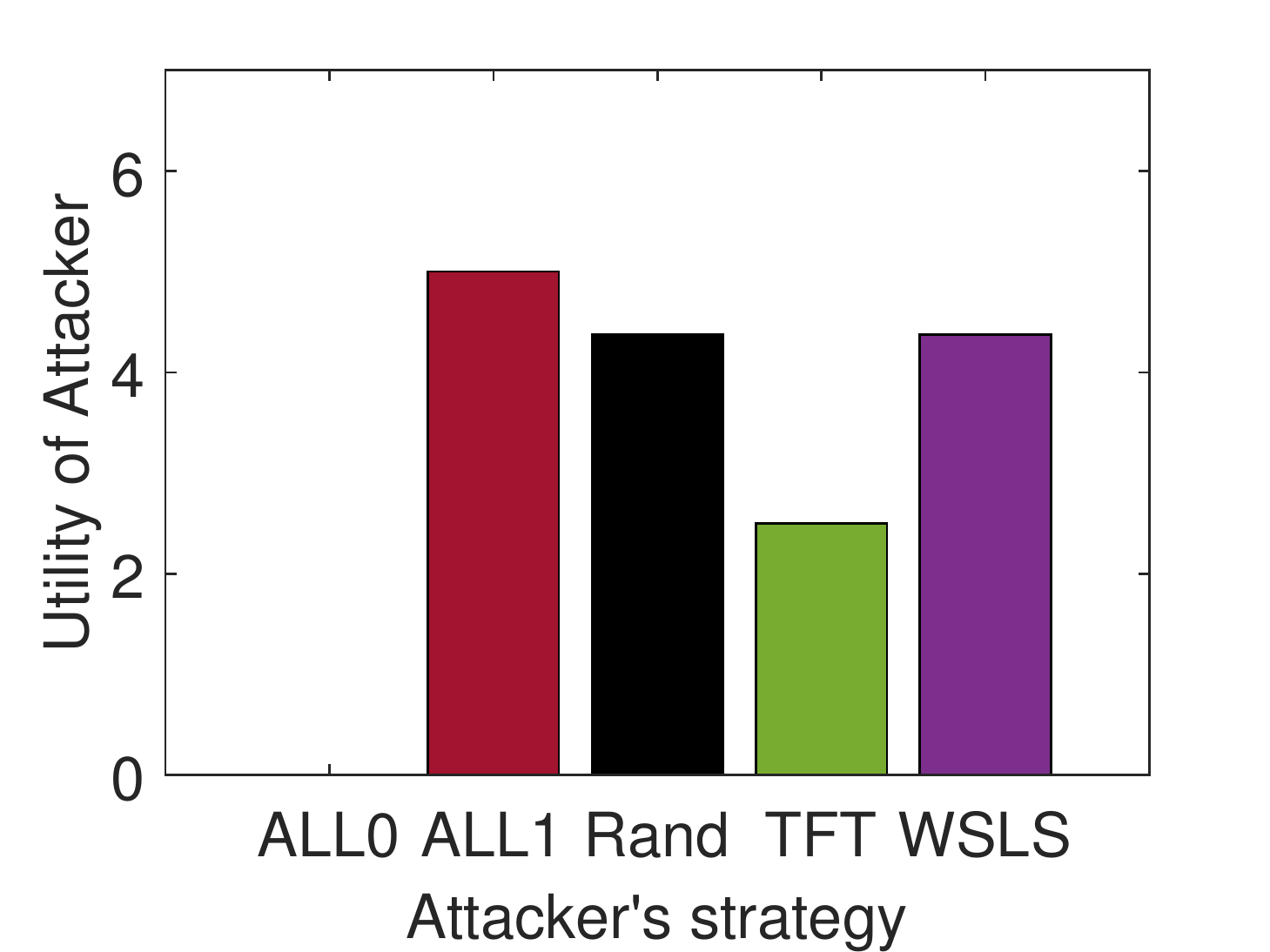}}
    \subfigure[WSLS]{
    \label{fig:subfig:5f}
    \includegraphics[scale=0.286]{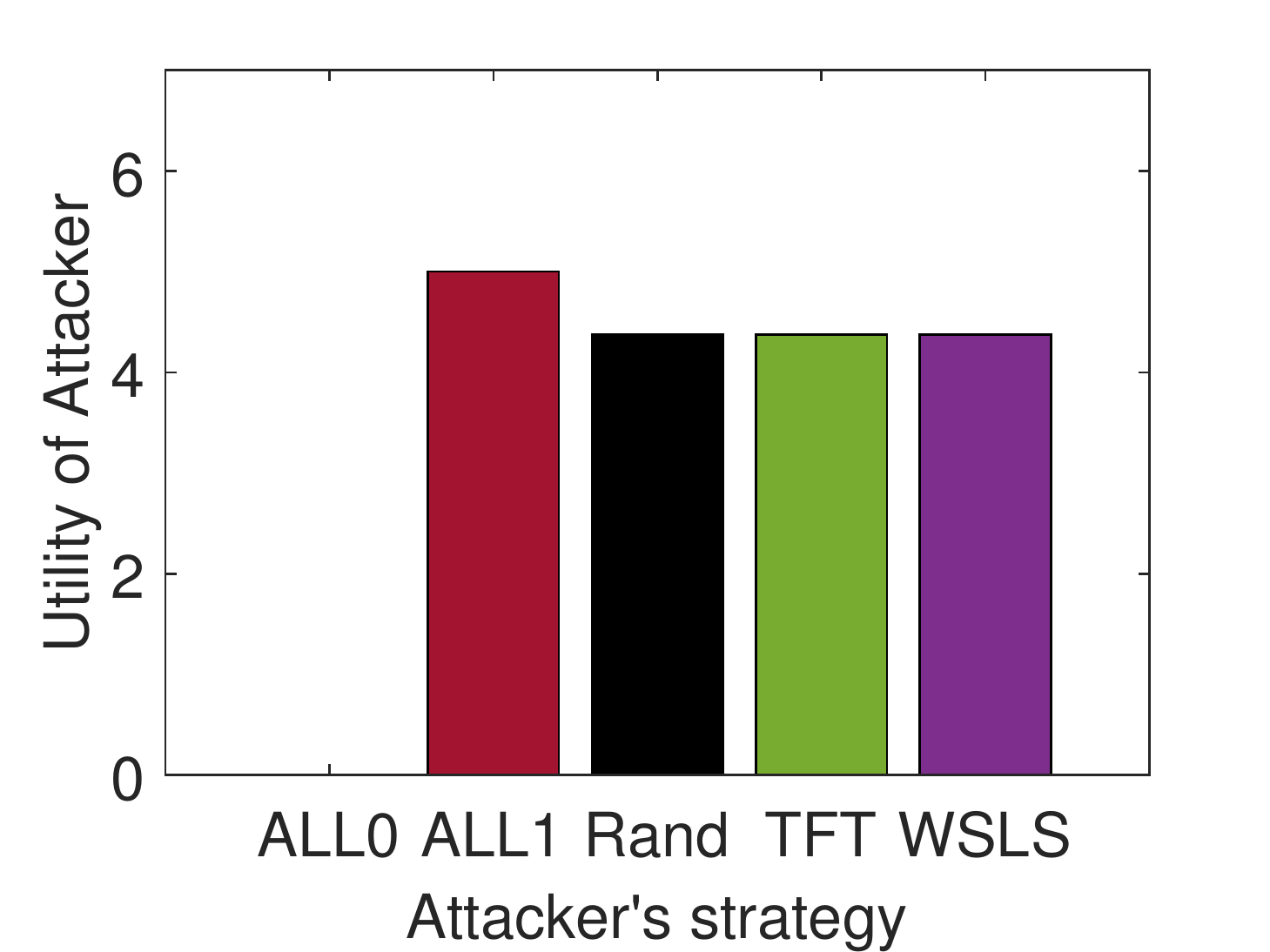}}
\caption{The attacker's utility under different strategy combinations of the attacker and the defender in the deterministic model.}
  \label{fig:strategies} 
\end{figure}

By comparing Fig. \ref{fig:strategies}(a) with the other five figures, we can easily find that when the attacker adopts ALL1, Rand, TFT and WSLS strategies, the defender's ZD strategy can effectively control the attacker's utility at a lower level. This can prove that unless the attacker adopts the ALL0 strategy, the ZD strategy is better than other classic strategies. However, in the real audit environment, it is almost impossible for the attacker to adopt ALL0 strategy, because it means that the attacker does not attack at all. Similarly, we can find in Fig. \ref{fig:strategies}(b) that if the defender adopts the ALL0 strategy, she can achieve good results in some cases but the rest can be bad, which reflects that the inactive defender suffers heavy losses when the attacker attacks and can only hope that the attacker would quit, which can not happen in reality.

\begin{figure}[h]
\centering
  \centering
  \subfigure[Attacker's strategy: ALL0]{
    \label{fig:rocall0}
    \includegraphics[scale=0.265]{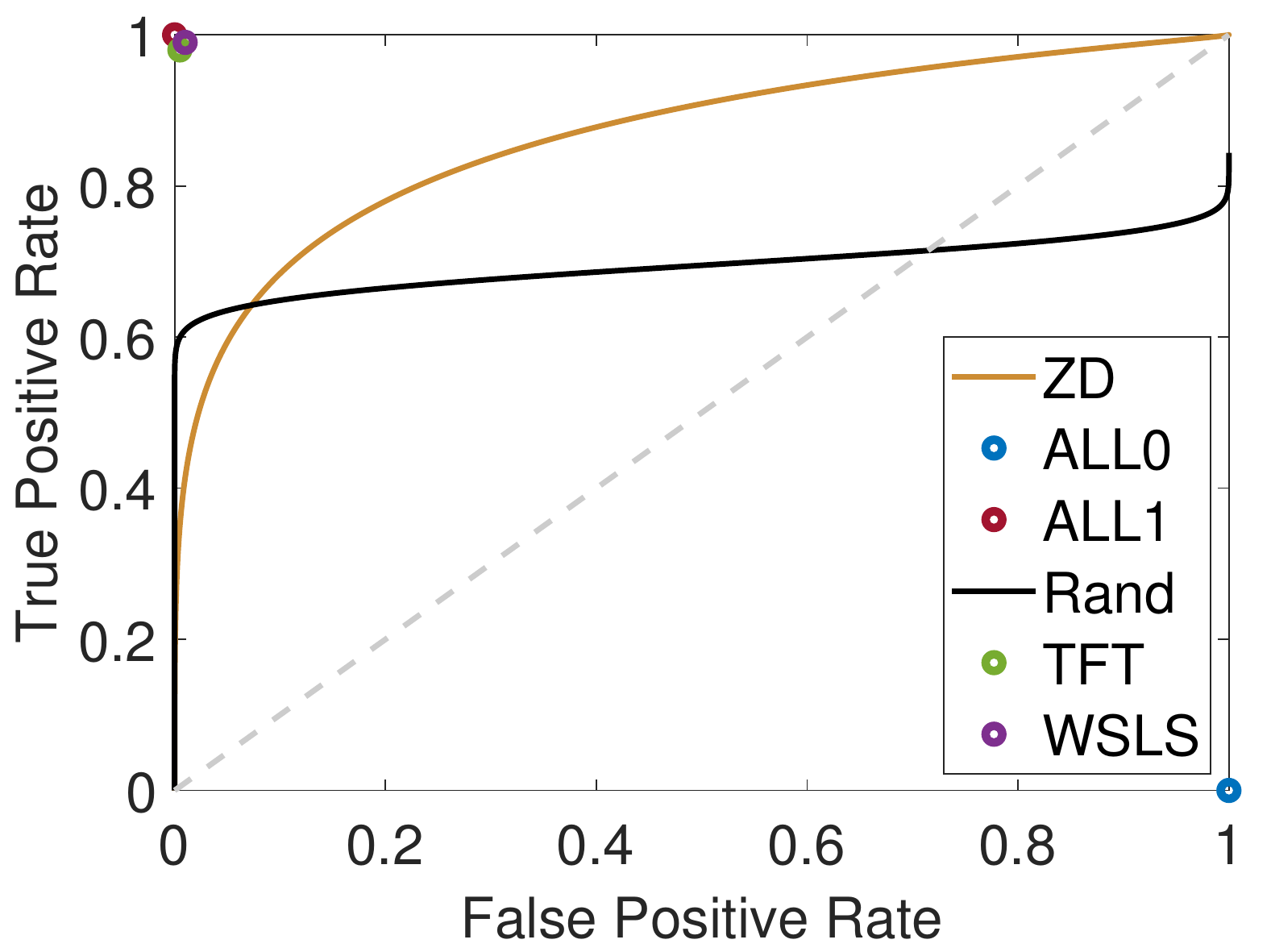}}
  \subfigure[Attacker's strategy: ALL1]{
    \label{fig:rocall1}
    \includegraphics[scale=0.265]{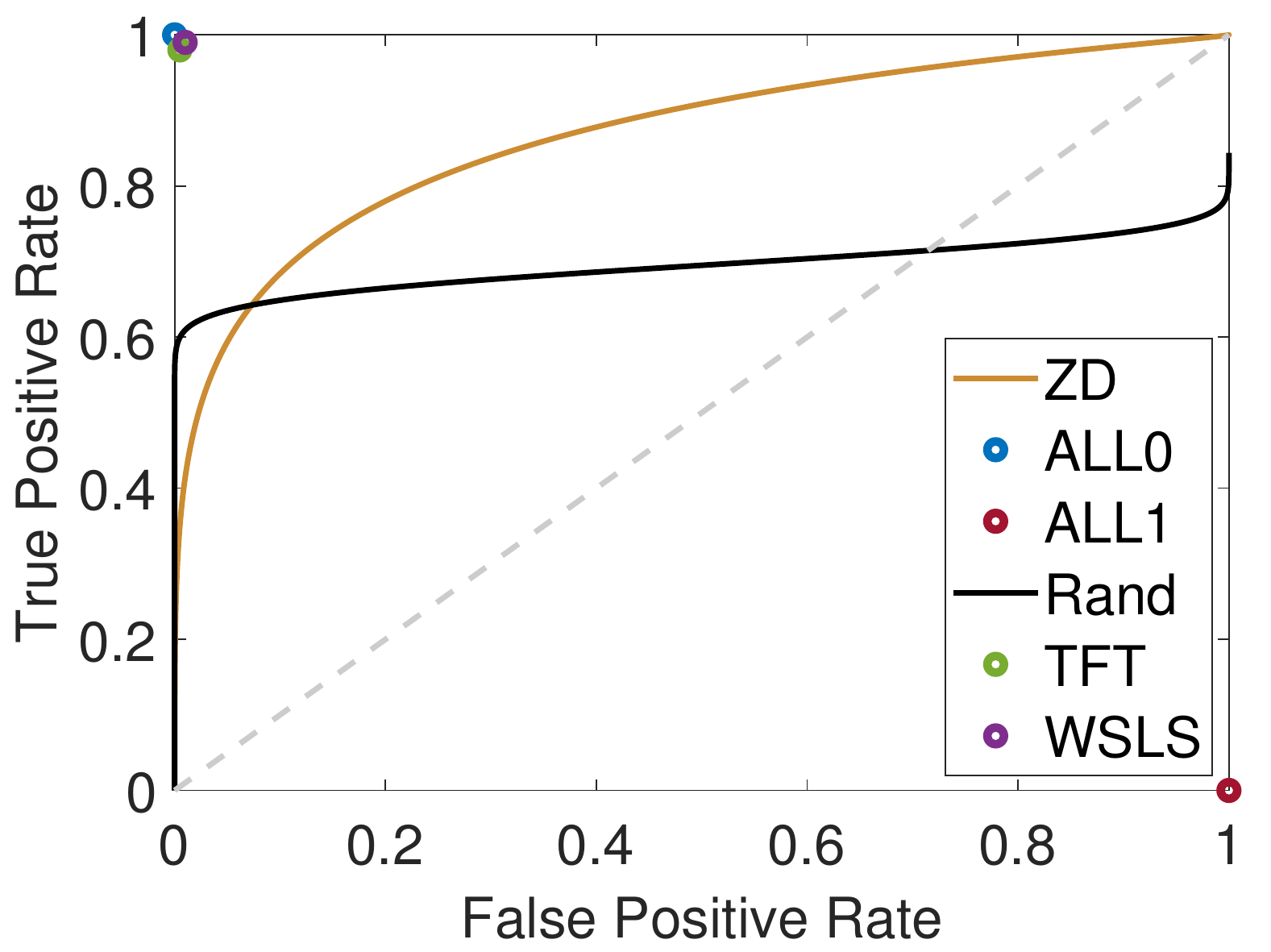}}
  \subfigure[Attacker's strategy: Rand]{
    \label{fig:rocrand}
    \includegraphics[scale=0.265]{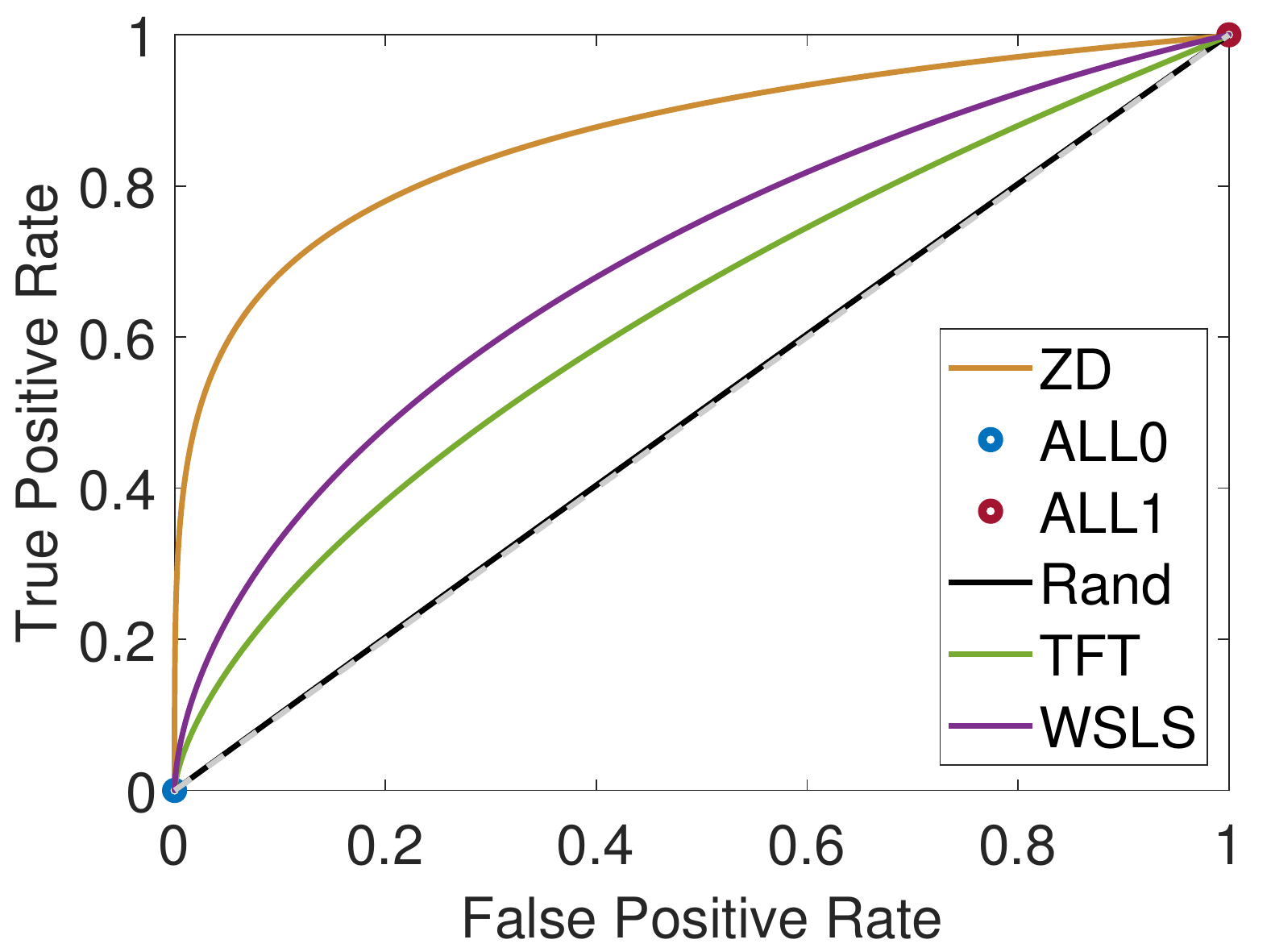}}
  \subfigure[Attacker's strategy: TFT]{
    \label{fig:roctft}
    \includegraphics[scale=0.265]{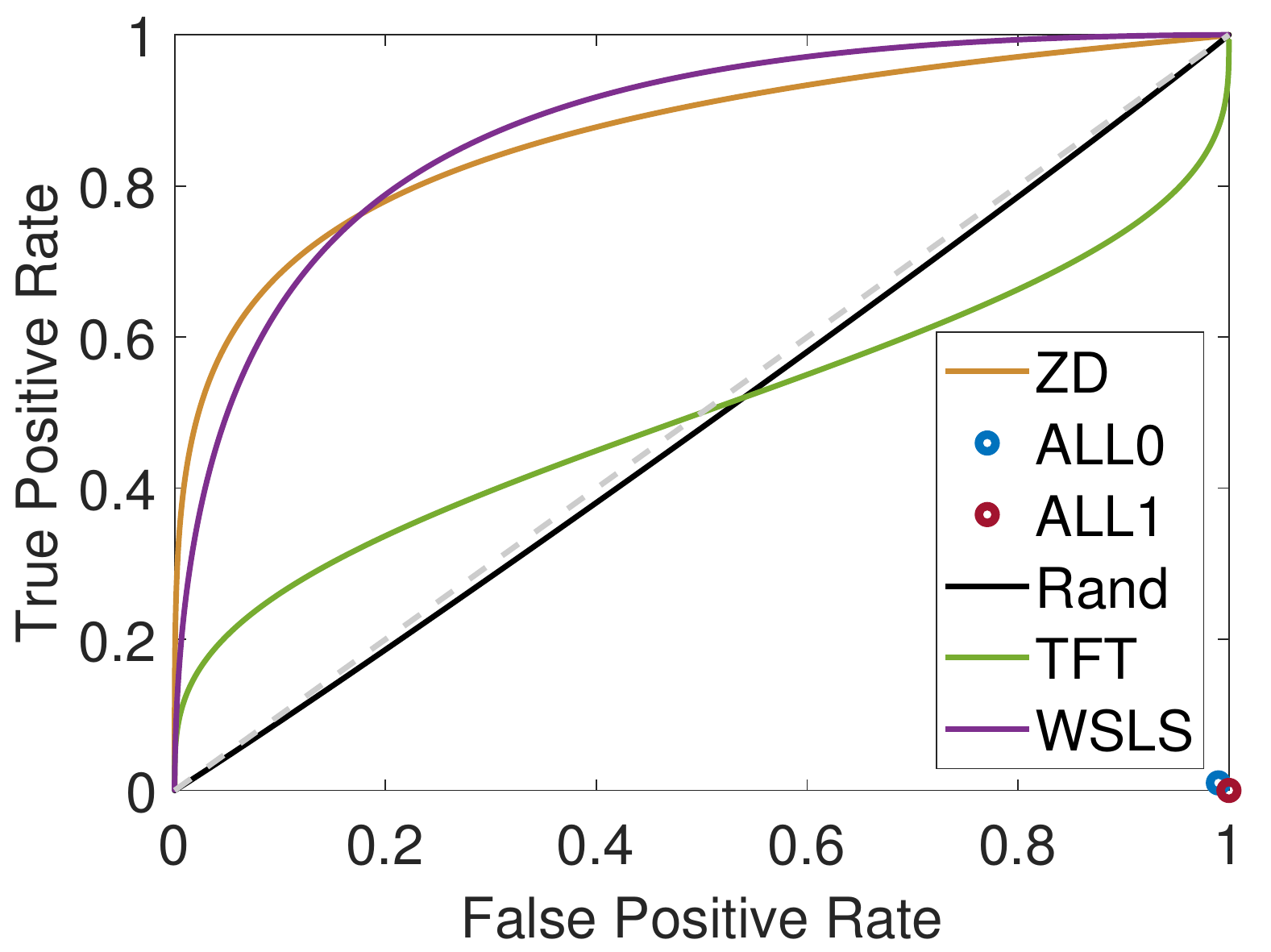}}
  \subfigure[Attacker's strategy: WSLS]{
    \label{fig:rocwsls}
    \includegraphics[scale=0.265]{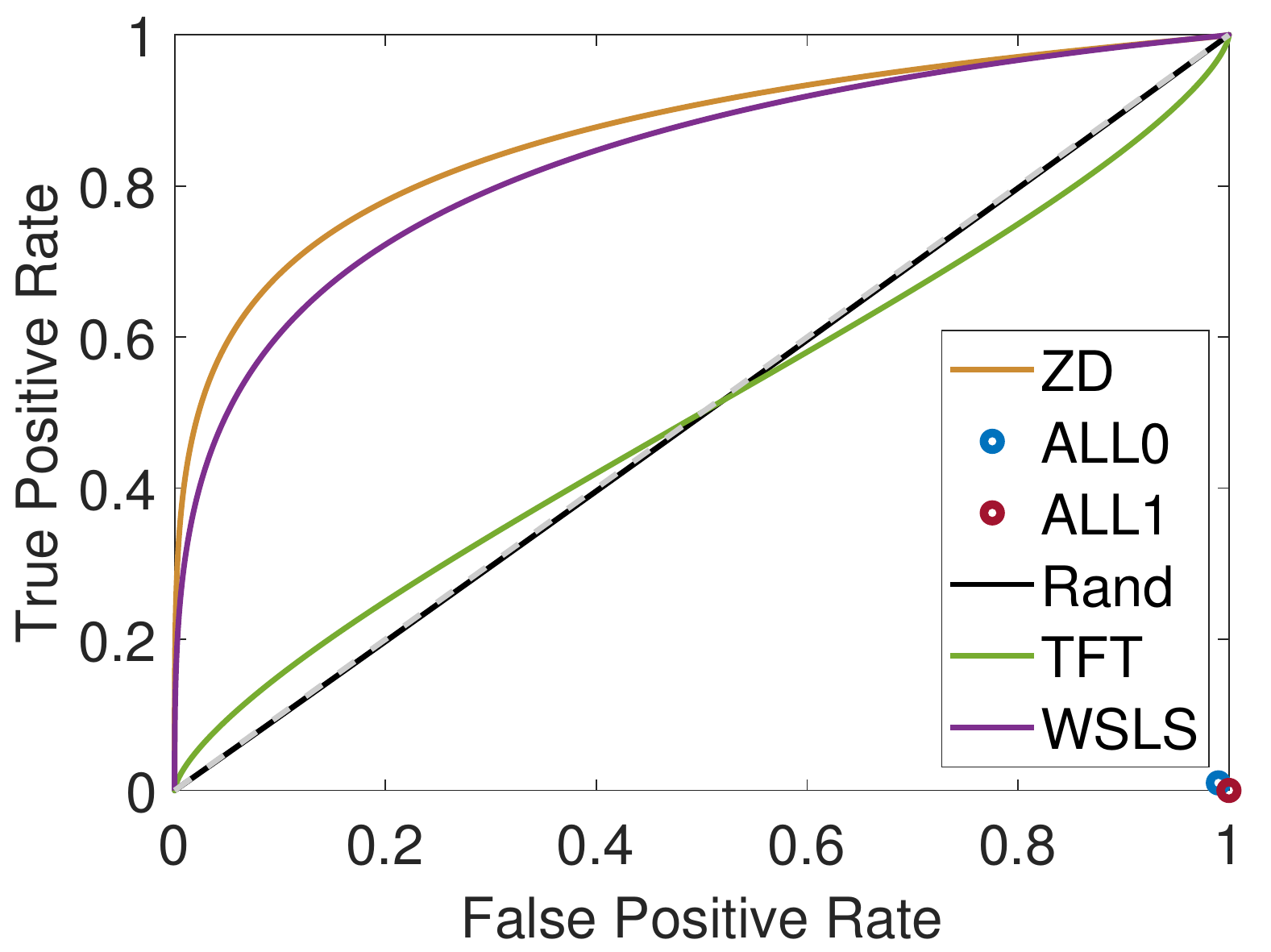}}
\caption{ROC curves under different strategy combinations of the defender and the attacker in the deterministic model.}
  \label{fig:ROC} 
\end{figure}

Next, we explore the detection performance of the ZD strategy for potential attacks and plot the Receiver Operating Characteristic (ROC) curves for the defender deploying the ZD strategy and other strategies when the attacker uses a classic strategy in Fig. \ref{fig:ROC}. We regard a test sample as a true positive if the defender chooses to send a signal and the attacker chooses to attack. Similarly, we define a sample as a false positive when the defender sends a signal while the attacker does not attack. Assuming a sample as a true negative if the defender does not send any signal but the attacker carries out the malicious action, and as a false negative if both sides do nothing. The x-axis depicts the False Positive Rate (FPR). Denoting FP and TN as the numbers of false positive samples and true negative samples, respectively, we can calculate $\text{FPR}=\frac{\text{FP}}{\text{FP}+\text{TN}}$. The y-axis represents the True Positive Rate (TPR), which is calculated by $\text{TPR}=\frac{\text{TP}}{\text{TP}+\text{FN}}$ with TP denoting the number of true positive samples and FN denoting the number of false negative samples.

From Fig. \ref{fig:ROC}, we can see that the ZD strategy outperforms almost all other strategies since its Area Under the Curve (AUC) is larger than the AUCs of other strategies. Besides, the gray dotted line represents the ROC curve of random guessing with AUC$=0.5$, which is used as a reference for comparison. Specifically, in Figs. \ref{fig:ROC}(a) and (b), the ROC curves of WSLS, TFT, ALL1, and ALL0 strategies degenerate to the point (0,1) or (1,0) as the defender only executes the same action when the attacker deploys the ALL0 or ALL1 strategy. For example, when the attacker uses the ALL1 strategy and the defender adopts the TFT strategy, the action in each round is $ad=(1,1)$, refering to the point (0,1) in the ROC curve. In Fig. \ref{fig:ROC}(d), the AUC of the WSLS strategy is close to that of the ZD strategy, which indicates that the performance of these two strategies is quite similar when the attacker adopts the TFT strategy.

\subsection{Maximizing the Utility Difference using the ZD Strategy}

\begin{figure}
  \centering
  \subfigure[]{
    \label{fig:subfig:7a} 
    \includegraphics[scale=0.286]{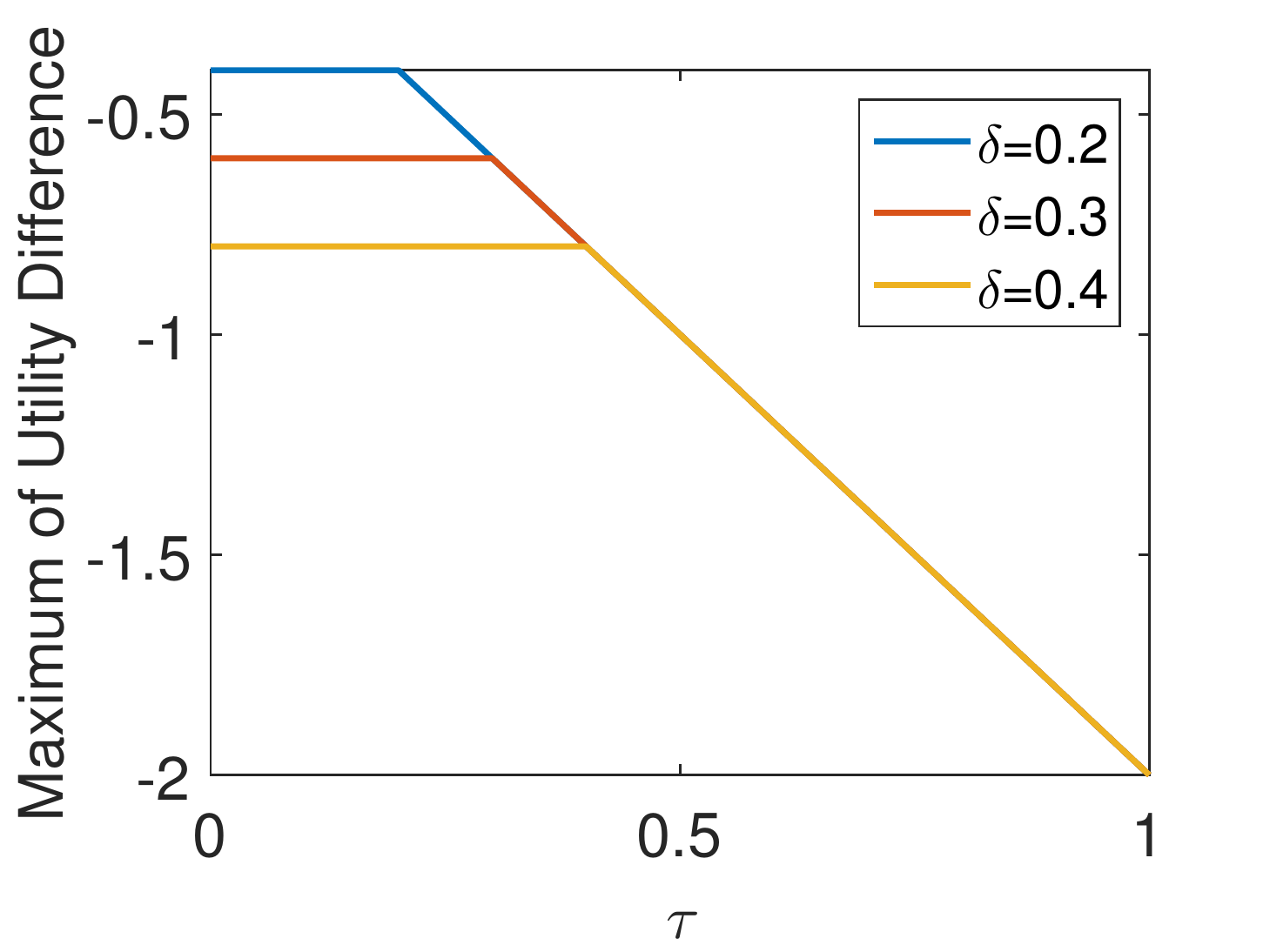}}
  \subfigure[]{
    \label{fig:subfig:7b} 
    \includegraphics[scale=0.286]{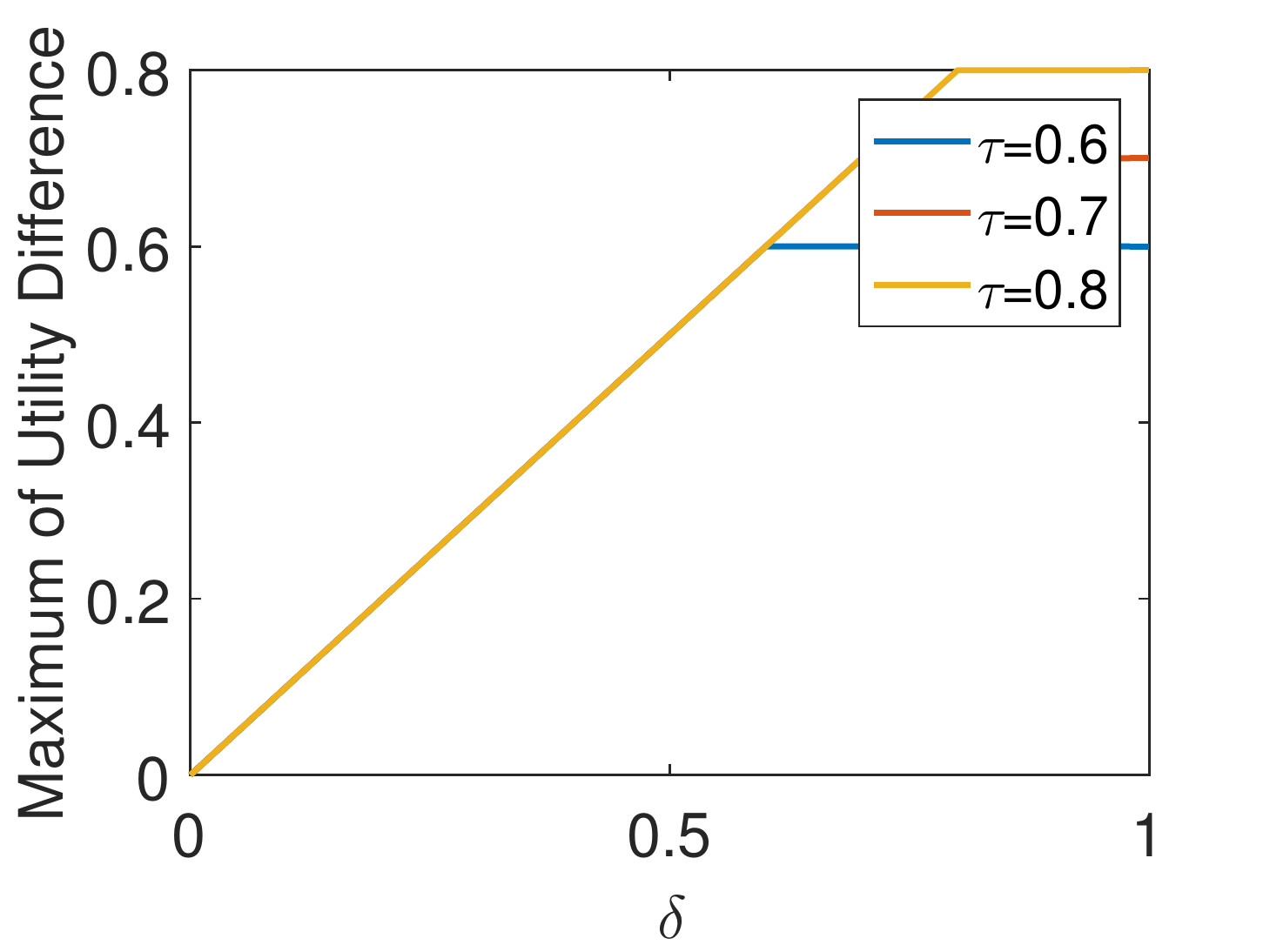}}
  \subfigure[]{
    \label{fig:subfig:7c} 
    \includegraphics[scale=0.286]{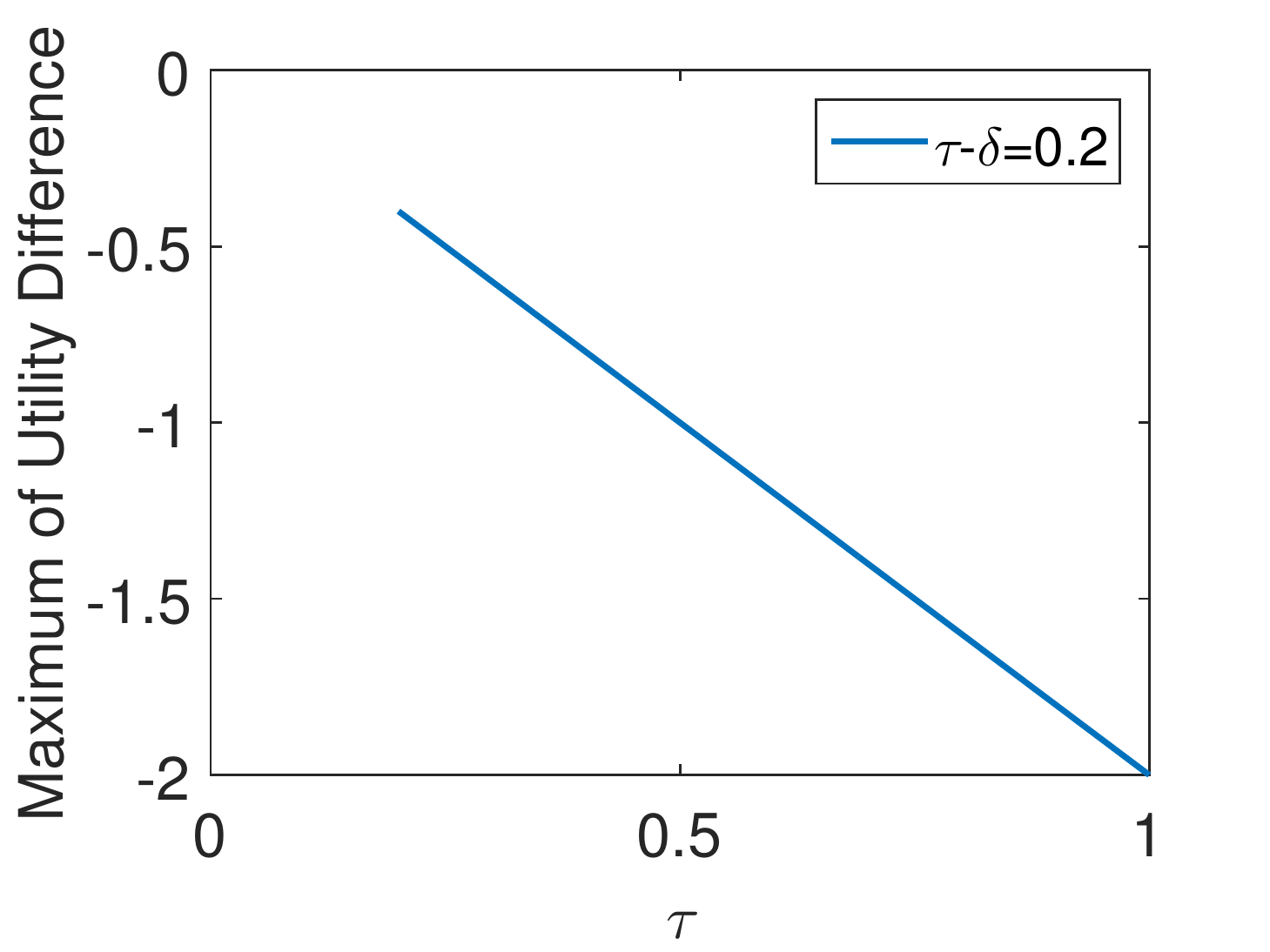}}
  \subfigure[]{
    \label{fig:subfig:7d} 
    \includegraphics[scale=0.286]{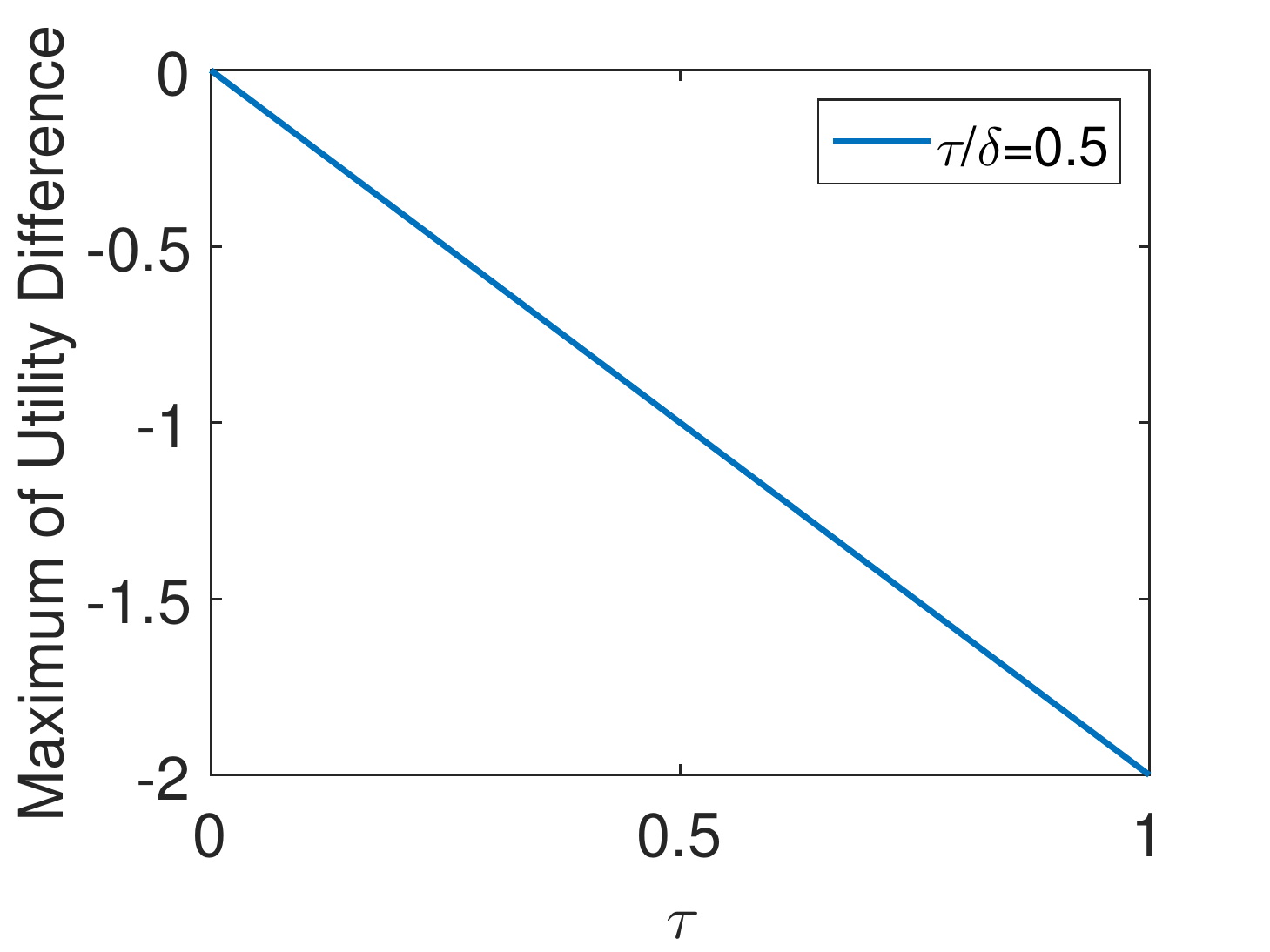}}
  \caption{The optimization goal $\tilde{u}_d-\tilde{u}_a$ changes with parameters combination in the probabilistic model.}
  \label{fig:optcorrectness} 
\end{figure}

We investigate the correctness and effectiveness of our proposed strategy for optimizing the utility difference between the defender and the attacker. In this part, we mainly show the experimental results of the probabilistic model. As demonstrated in Section \ref{Section5}, it is easy to draw similar conclusions in the probabilistic model and the deterministic model by setting $\tau=1$ and $\delta=0$. In Fig. \ref{fig:optcorrectness}, we present the solution of the optimization problem that maximizes $\tilde{u}_d-\tilde{u}_a$. Based on the parameter setting mentioned before, we found that there is a feasible solution for $p_i$ in the constraint condition if and only if $\phi<0$. We plot the figure of the maximized utility difference changing with $\tau$ and $\delta$ when $\phi=-1$. It can be seen that under this condition, the maximum value of the optimization target is negatively correlated with $\tau$ and positively correlated with $\delta$. This is because if the defender considers the utilities of both herself and the attacker, she has to consider appropriately reducing the probability of auditing ($\tau$) after signaling because of the cost of the audit. But the defender cannot reduce this probability without any limit, because when it reaches a certain value, it no longer has an impact on the maximum value of the optimization goal. Fig. \ref{fig:optcorrectness}(c) shows that if the defender changes $\tau$ and $\delta$ at the same time while keeping the difference between them unchanged, $\tau$ will have a linear effect on the maximum utility difference. It is worth noting that in Fig. \ref{fig:optcorrectness}(d) if we set $\tau$ and $\delta$ proportionally, the influence of $\tau$ on the optimization goal is also linear. 

\begin{figure}[h]
\centering
  \centering
  \subfigure[ZD]{
    \label{fig:subfig:8a}
    \includegraphics[scale=0.286]{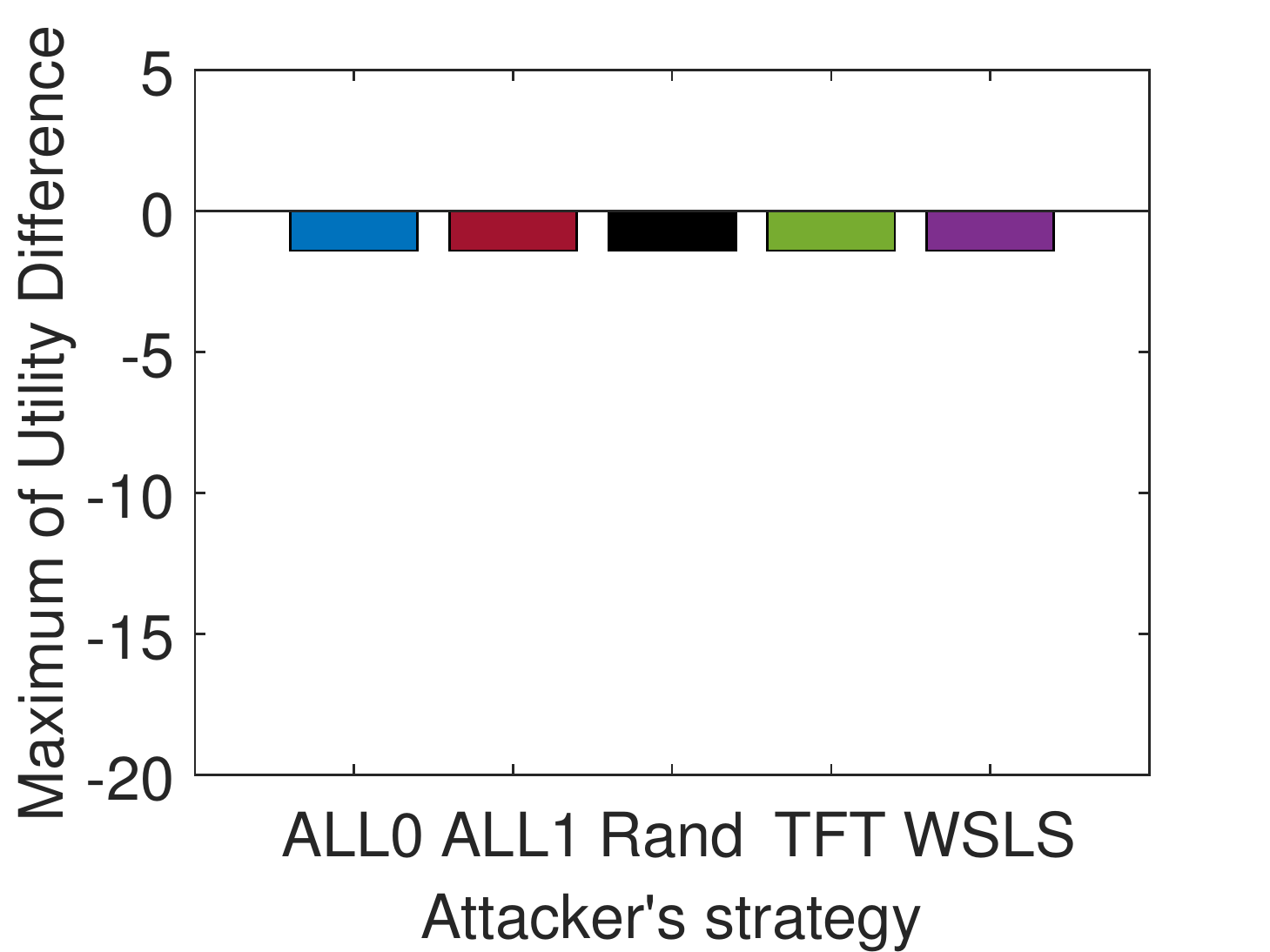}}
  \subfigure[ALL0]{
    \label{fig:subfig:8b}
    \includegraphics[scale=0.286]{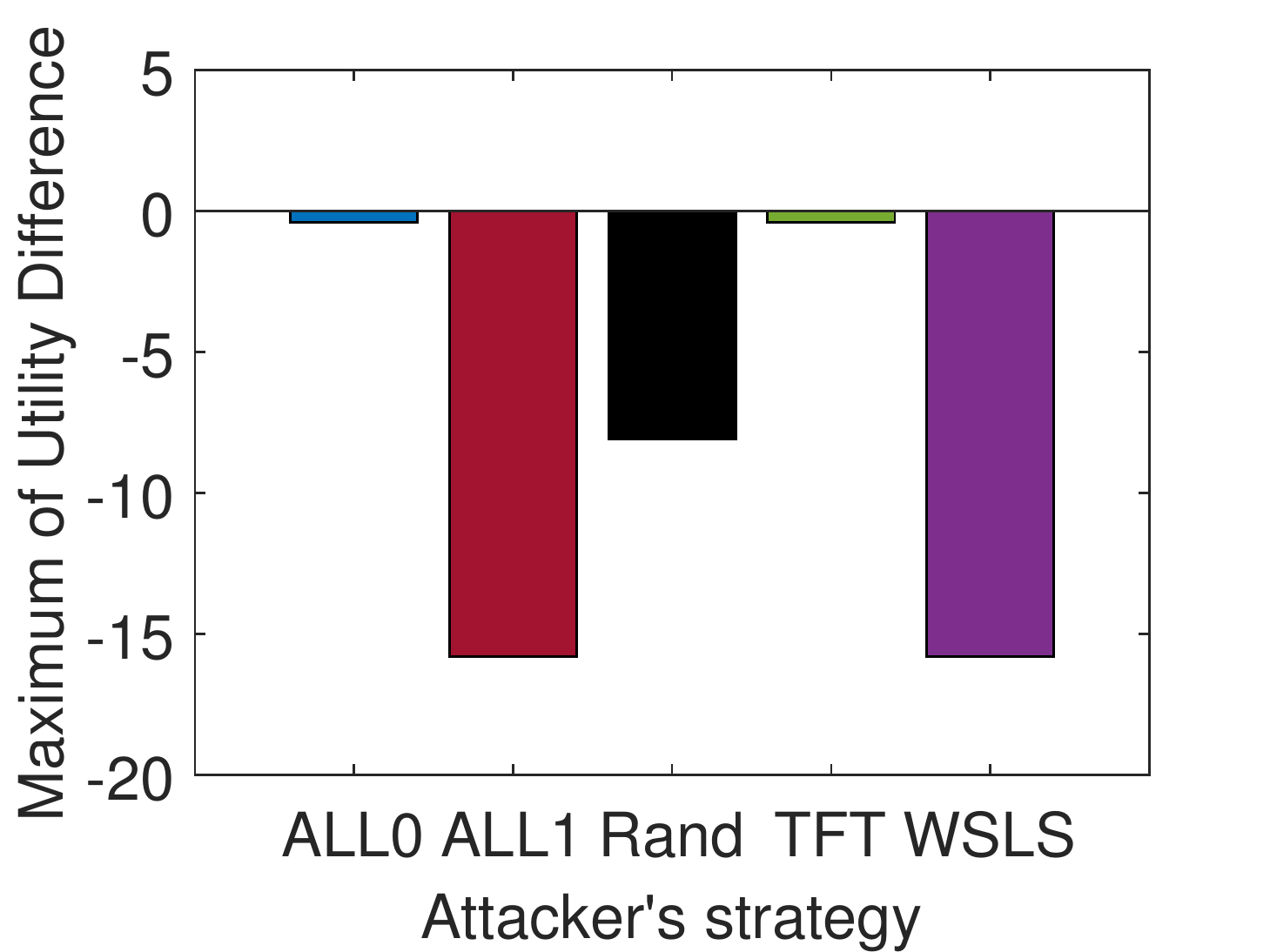}}
  \subfigure[ALL1]{
    \label{fig:subfig:8c}
    \includegraphics[scale=0.286]{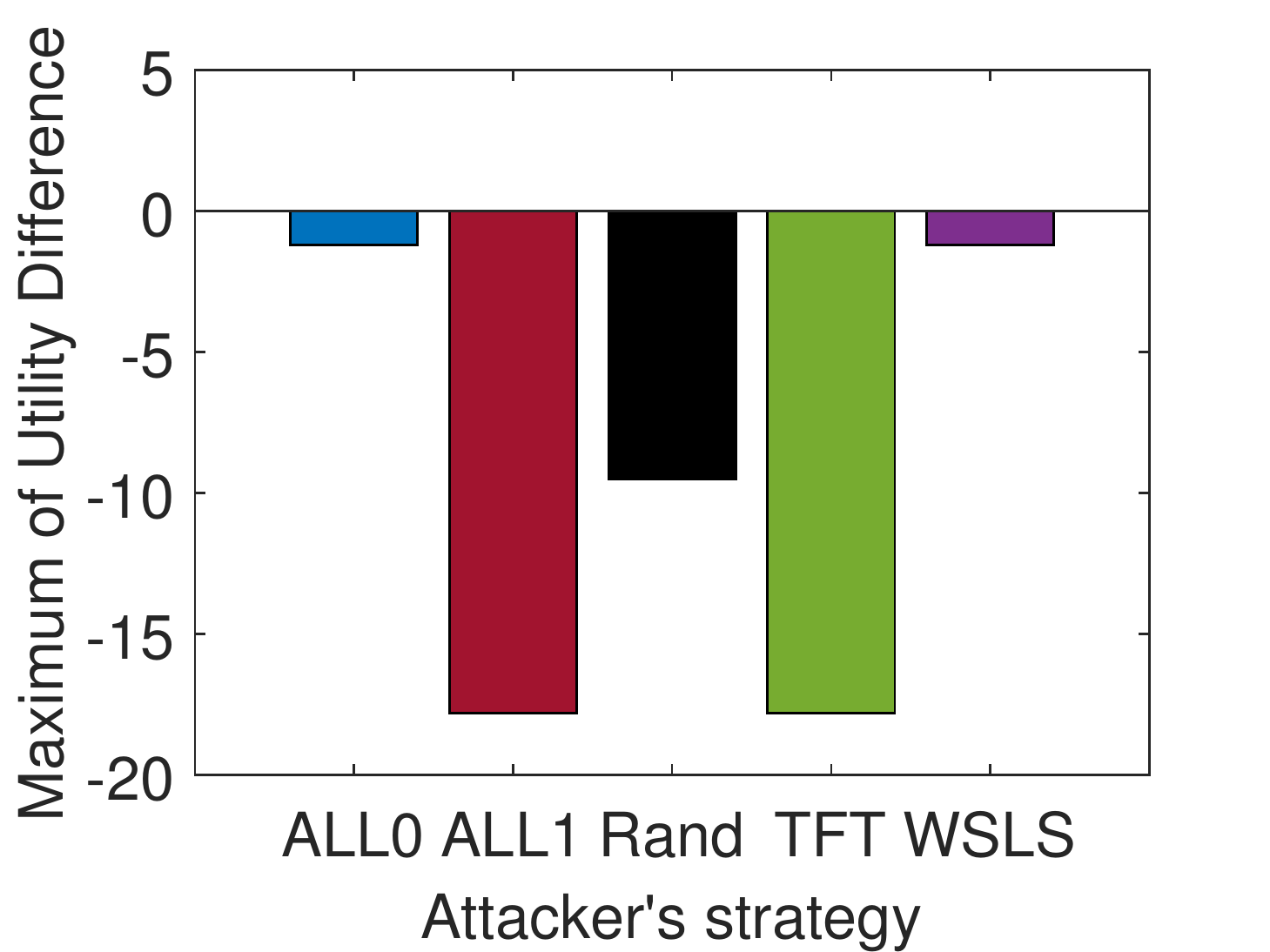}}
  \subfigure[Rand]{
    \label{fig:subfig:8d}
    \includegraphics[scale=0.286]{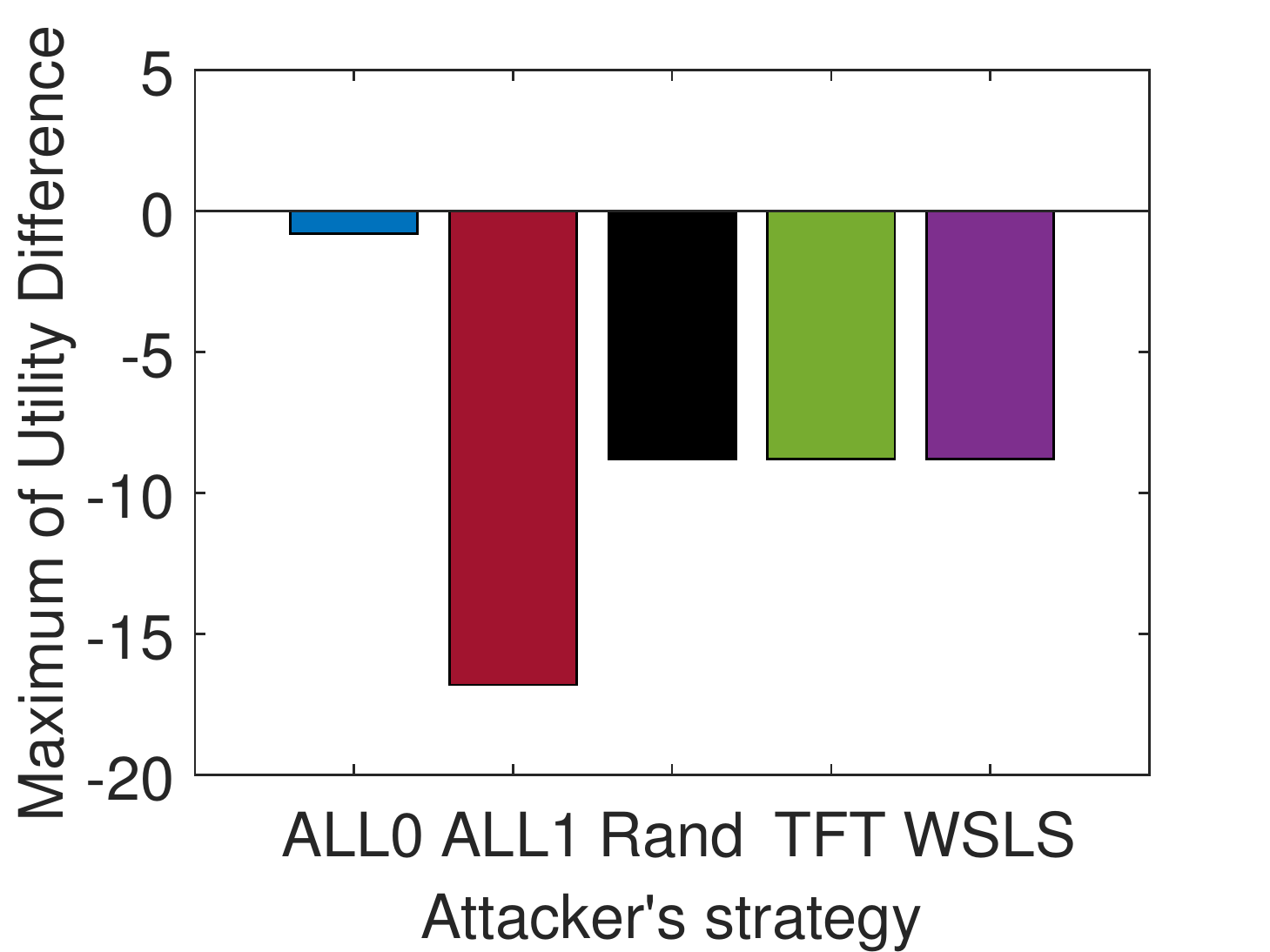}}
  \subfigure[TFT]{
    \label{fig:subfig:8e}
    \includegraphics[scale=0.286]{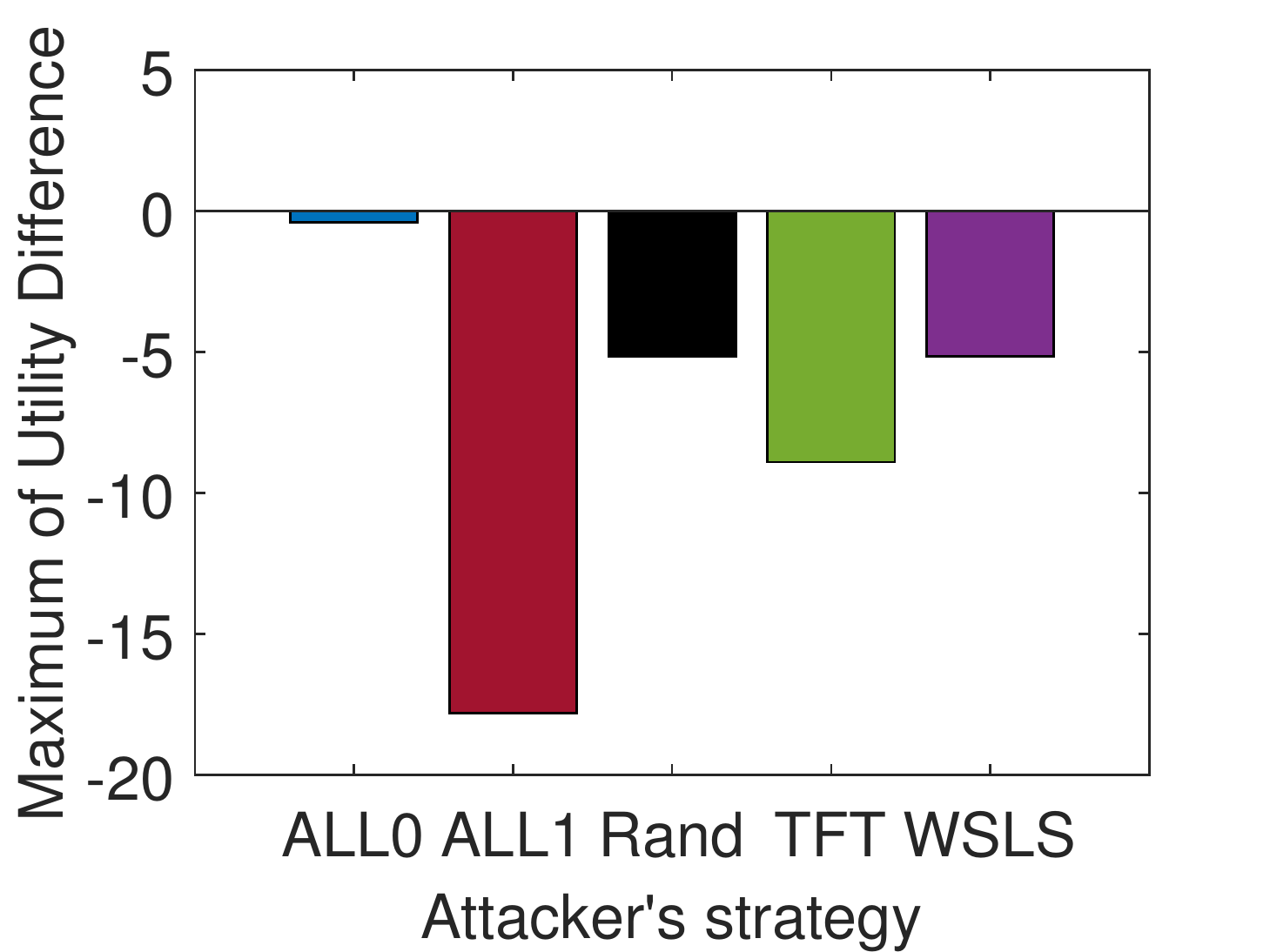}}
    \subfigure[WSLS]{
    \label{fig:subfig:8f}
    \includegraphics[scale=0.286]{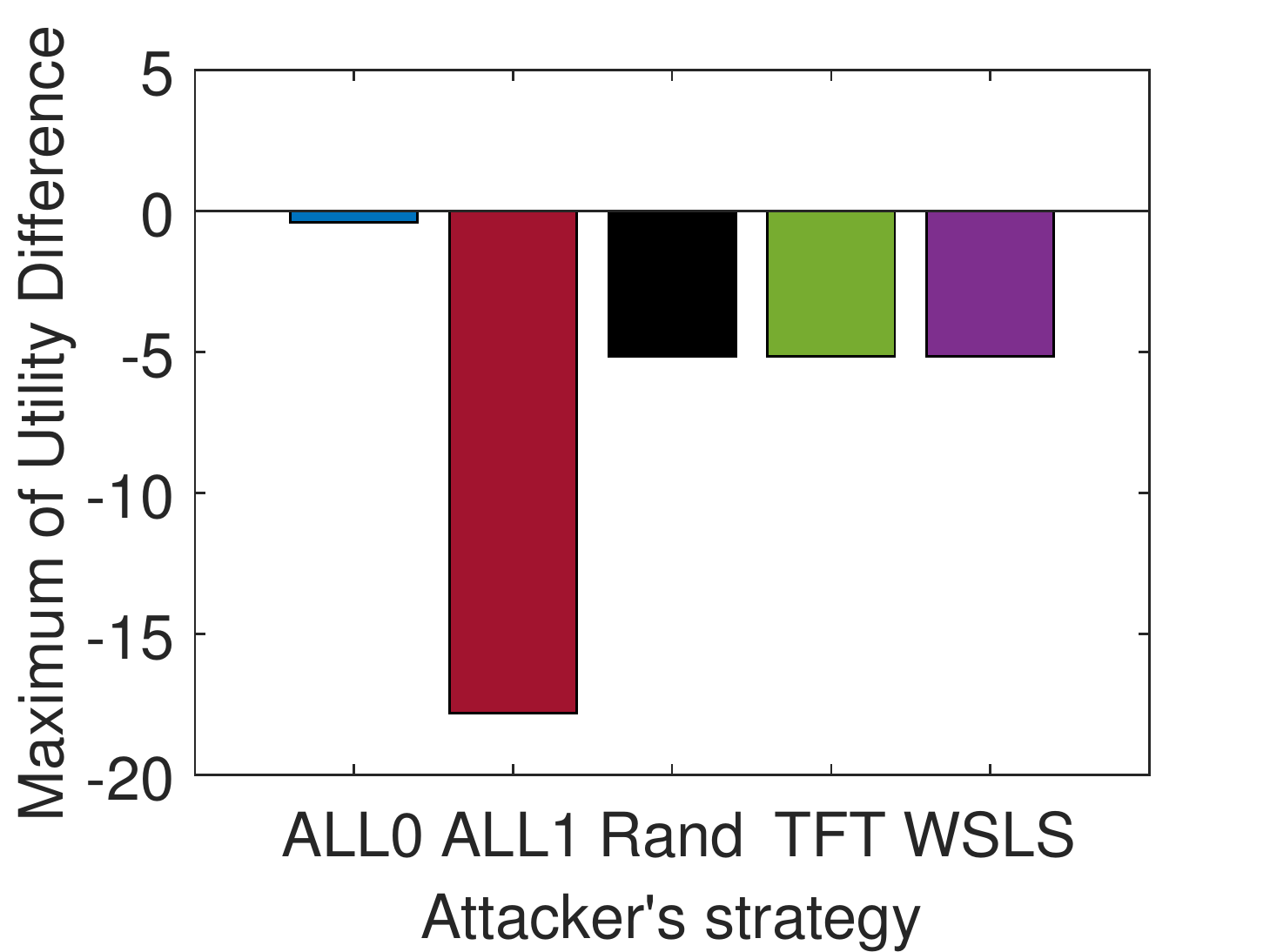}}
\caption{Maximum of $\tilde{u}_d-\tilde{u}_a$ under different strategy combinations of the attacker and the defender in the probabilistic model.}
    \label{fig:opteff} 
\end{figure}

To verify the effectiveness of our ZD strategy-based scheme, we set $\tau=0.6$ and $\delta=0.2$, and compare the optimization goal of ZD scheme with those obtained by other classic strategies, i.e., ALL1, Rand, TFT, and WSLS strategies. Fig. \ref{fig:opteff} displays the optimization goal $\tilde{u}_d-\tilde{u}_a$ when the defender takes different strategies. By comparing Fig. \ref{fig:opteff}(a) with the other five figures, one can conclude that the ZD strategy gets a larger maximum value of the optimization target, except for the situation that the attacker adopts the ALL0 strategy and some situations that the defender adopts the ALL0 strategy. However, it is rare for an attacker to adopt the ALL0 strategy. In this case, an active defender consumes more audit budget than an inactive defender, which makes the total utility less. Besides, if the defender adopts the ALL0 strategy for a long time, then she can only hope that the attacker will never attack (also adopts the ALL0 strategy), which hardly occurs in actual situations. So in most common cases, using the ZD strategy can effectively make the difference between the defender's utility and attacker's utility stay at a high level.

\section{Results and Discussion}
In the previous sections, we present a deterministic model and a probabilistic model to describe the sequential games in the signaling-based audit mechanism. With the help of the extended ZD strategy, we can enable the defender to unilaterally control the attacker's utility and maximize the utility difference between the defender and the attacker. However, the following limitations remain in our model assumptions and experimental design.

\begin{itemize}
\item \textbf{What if the attacker uses the ZD strategy?}
In our experiments, we display the results of the defender's ZD strategy playing against other strategies of the attacker. However, we do not consider what would happen if the attacker also uses the ZD strategy. This may happen in an actual situation since the ZD strategy is powerful.
\item \textbf{The assumption of same values for all data.}
The data might have different sensitivities, reflecting different importances. Therefore, if the data protected by the defender have different values in real situations, the proposed ZD strategy and its control capability would change, which is one of the limitations of the current assumptions. 
\item \textbf{Utility maximization of the defender.}
Our proposed ZD strategy can achieve robust control over the attacker's utility and maximize the utility difference. However, it is not clear whether it would still work when only the defender's utility is required to be maximized.
\end{itemize}

\section{Conclusion and Future Work}
In this paper, we propose two sequential game models to describe the interaction between the defender and the attacker, where the auditing behavior of the defender is deterministic and probabilistic. Using the ZD strategy allows the defender to unilaterally control the attacker’s utility no matter what strategy the attacker uses. In addition, an optimization scheme is designed for the defender based on the ZD strategy to control the utility difference between the defender and the attacker. Via comparing the ZD strategy with other classic strategies, experimental results show that the ZD strategy has better performance in controlling the attacker's utility as well as maximizing the utility difference between the defender and the attacker. 

In the future, we will study the situation where the attacker also adopts the ZD strategy and consider how the defender can make better defensive actions. We are going to further consider the implementation and practicality of the ZD strategy for the audit game when the stored data have different values. Moreover, we intend to design a new strategy to maximize the defender's utility. 

\bibliographystyle{./IEEEtran}
\bibliography{./reference.bib,./IEEEexample}


\end{document}